\documentclass[letterpaper]{JHEP3}
\usepackage{amssymb,amsmath,epsfig,cite}

\def\abar{\overline{a}}
\def\bbar{\overline{b}}
\def\kbar{\overline{k}}
\def\pbar{\overline{\p}}
\def\Pbar{\overline{P}}

\def\ubar{\overline{u}}
\def\vbar{\overline{v}}
\def\wbar{\overline{w}}
\def\zbar{\overline{z}}
\def\xibar{\overline{\xi}}
\def\chibar{\overline{\chi}}

\def\Qbar{\overline{Q}}

\def\p{\partial}
\def\Re{{\rm Re}}

\def\a{\alpha}

\def\G{\Gamma}
\def\d{\delta}
\def\D{\Delta}

\def\th{\theta}

\def\ka{\kappa}
\def\la{\lambda}
\def\La{\Lambda}
\def\s{\sigma}
\def\t{\tau}
\def\w{\omega}

\def\({\left (}
\def\){\right )}
\def\[{\left[}
\def\]{\right]}
\def\goesto{\rightarrow}

\def\EV#1{\left\langle#1\right\rangle}

\def\sgn{{\rm sgn}}
\def\ha{{1\over2}}

\def\Tr{{\rm Tr}}

\preprint{arXiv.org/0806.0181 [hep-th] \\
PUPT-2267}

\title{
Holographic Double Diffractive Scattering 
}

\author{
Christopher P. Herzog$^{12}$,
Steve Paik$^1$, 
Matthew J. Strassler$^{13}$,
Ethan G. Thompson$^1$ \\
$^1$Department of Physics, University of Washington, Seattle, WA 98195--1560\\
$^2$Department of Physics, Princeton University, Princeton, NJ 08544\\
$^3$Department of Physics and Astronomy, Rutgers University, Piscataway, NJ 
08854--8019
}

\abstract{
The holographic description of Pomeron exchange in a strongly-coupled
gauge theory with an AdS dual is extended to the case of two to three
scattering.  We study the production event of a central particle via
hadron-hadron scattering in the double Regge kinematic regime of 
large center-of-momentum energy and fixed momentum transfer.
The computation reduces to the overlap of a holographic wave
function for the central particle with a source function for the Pomerons. The
formalism is applied to scalar glueball production and the resulting
amplitude is studied in various kinematic limits.
}

\begin{document}

\section{Introduction}

Regge phenomena have been studied, both in experiments and in
theoretical contexts, for several decades.  Experiments investigating
hadron scattering, and associated theoretical attempts to understand
the data, led in the 1960s to a number of important developments that
continue to play a role in current research.  The general Regge
approach to writing amplitudes as complex-valued functions of
Mandelstam variables, and investigating them using Mellin transforms,
had some general success in characterizing observed amplitudes that
appeared at the time to be inconsistent with quantum field theory.
The development of ``dual resonance'' models that captured various
features of scattering amplitudes, including $s$-$t$ channel duality,
towers of resonances lying on Regge trajectories $\alpha(m^2)$ for
$m^2>0$, and near-forward scattering amplitudes proportional to
$s^{\alpha(t)}$ for $t<0$, were the first steps along the road to a
consistent string theory.  Of course, hadronic scattering experiments
and standard string theory diverged in large-angle scattering (where
the amplitudes for the former fall as powers of $s$, while those of
the latter fall exponentially) and in deep inelastic scattering (where
Bjorken scaling suggested hard, weakly-interacting substructure).  The
theory of QCD supplanted string theory as the leading model for strong
interactions, and its success is spectacular.  Further, it was soon
learned that consistent string theories needed to be defined with
massless spin-two particles, which QCD of course lacks.  Yet the fact
that Regge phenomena are evident in the data, in regimes
which are not amenable to study using
perturbative QCD, and the successes of the early phenomenogical (and
inconsistent) string theories, have left lingering questions about the 
relationship between string theory and the physics of hadrons.

A better understanding of the relationship between QCD and string
theory has now emerged from the discovery of an apparent (but still
unproven) duality between non-abelian gauge theory and string theory
in curved spaces.  This duality conjecture \cite{adscft} contends that
four-dimensional gauge theories which are asymptotically conformal are
identical to ten-dimensional string theories propagating on spaces
that are asymptotically five-dimensional anti-de Sitter space
($AdS_5$) times a compact five-manifold $W$.  The conjecture has
provided a setting in which the successes and failures of the old
string theories could be substantially clarified.  The inability of
four-dimensional flat-space string theories to match general features
of high-energy QCD amplitudes was shown to be rectified in the
scattering of strings propagating on the appropriate curved spaces;
exponential fall-off with $s$ in large-angle scattering is replaced
with power laws \cite{HS}.  Bjorken scaling does not hold, but
its failure is replaced \cite{deepinelastic} with a more generalized scaling
discussed by Kogut and Susskind in the context of conformal field
theory \cite{Kogut:1974ni}.  The Regge phenomena in string theory
amplitudes manage to reproduce diverse phenomena of gauge theory in
various Regge-like regimes \cite{HS,Brower:2006ea}.  In high-energy
scattering at small momentum transfer $|t|$, where experiments show
the classic Regge forward peak, the string theory computation reduces
approximately to four-dimensional flat-space string theory, and the
Regge peak in the string theory is transferred directly to the gauge
theory.  At larger negative $t$, however, and in some other regimes as
well, the fifth dimension of the asymptotically-$AdS_5$ space begins
to play a role, washing out the stringy Regge behavior, and instead
reproducing phenomena that mirror the amplitudes calculated by
Balitsky, Fadin, Kuraev and Lipatov (BFKL) \cite{BFKL} in
gauge theory, as well as other phenomena.  At the same time, the
string theory also provides a discrete set of states lying on a set of
Regge trajectories.  Thus many of the essential features of gauge
theory, especially those which are not found in purely perturbative
QCD, have been shown to be reproduced in this ``gauge/string
correspondence.''

In this paper, we continue the process of learning about Regge physics
in contexts that are challenging to understand fully in QCD.  All
previous studies of Regge physics in the gauge/string correspondence
have been of $2\to 2$ scattering, and here we will attempt to extend
this to important $2\to 3$ phenomena in the so-called
``double diffractive'' or ``double Regge'' regime.  In this regime,
one considers $A B \to A B X$, where $A, B$ are hadrons that are
scattered at small angles without internal excitation or other
disruption, and $X$ is an object produced with a rapidity that lies
well between the rapidities of the outgoing $A$ and $B$.  Interesting
examples in this class that are more or less addressable
experimentally include cases where $A$ is a proton, $B$ is a proton or
antiproton, and $X$ is a glueball, a quarkonium state, or a Higgs
boson.  Diffractive quarkonium production has been observed at
Fermilab \cite{Affolder:2001nc}.  
Higgs production in double diffractive $pp$ scattering is
an important controversial topic, as experiments are being planned to
search for it at the LHC amid heated discussions about the expected
cross-section (see \cite{Forshawreview} for a brief review).  
We will however focus on something akin to $pp\to pp \
+$ glueball, as this is technically the easiest problem.  Our methods
will require some minor generalization for the study of quarkonium
production, and at least one additional technical advance for Higgs
production.

We should emphasize that our goals are strictly limited, yet
ambitious.  On the one hand, none of these processes in real-world QCD
can be reliably computed using the gauge/string correspondence.  The
correspondence allows detailed study of a certain class of gauge
theories, but unfortunately QCD is not among them.  No well-controlled
study of QCD dynamics, allowing for a computation of cross-sections
for comparison with experimental data, is currently possible.  While
some surprising success has been achieved modeling the spectrum of QCD, 
as well as some hadron couplings,
using five-dimensional effective field theories
\cite{erlichetal,daroldpomarol},  it
is quite another matter to compute scattering amplitudes in which
stringy phenomena play an essential role.  We know that some aspects
of QCD in scattering are not true in gauge theories for which stringy
computations are reliable.  For instance, Bjorken scaling is badly
violated, and Regge exponents differ by something of order unity from
those at weak coupling.

Our approach here is to treat those gauge theories for which the
gauge/string correspondence can be easily applied as {\it toy models}
for QCD.  We seek to identify phenomena which are universal or
quasi-universal in these toy models, and in turn, to understand the
degree to which they may apply also in all confining gauge theories,
including QCD.  Though the numerical details of our calculations will
not match QCD, we expect our toy model and QCD to share key
qualitative and semi-quantitative features.  Moreover, the
similarities and differences between our model and QCD should be
physically comprehensible.

The simplest possible use of our toy models would be a direct one:
it is possible that the behavior of our amplitudes as a function of
the kinematic variables may be similar to that of QCD.  A more subtle
but potentially more important use of these toy models is at the level
of general methodology and conceptual understanding.  Our goals here
are limited to the latter, and our results should be viewed as
exploratory, especially as data on the process computed below is
limited.  We hope this paper will be a useful step in the direction of
allowing computation of diffractive processes for which data is more
easily accessible.

\section{Preliminaries}

A gauge/string duality is a conjectured equivalence between a string
theory in an asymptotically hyperbolic geometry and a gauge theory in
a fewer number of dimensions.  The original and best established
example of this duality maps type IIB string theory in an $AdS_5
\times S^5$ background, where $AdS_5$ is five-dimensional anti-de
Sitter space and $S^5$ is a five-dimensional sphere, to the maximally
supersymmetric $SU(N)$ Yang-Mills theory in 3+1 dimensions.  This
supersymmetric field theory is scale invariant and is thus an example
of a conformal field theory (CFT).  As the hard-wall model we consider
in this paper is only a slight generalization of this original version
of the duality, it is worthwhile to review the details of the
$AdS_5/{\rm CFT}_4$ correspondence.

The original gauge/string duality can be motivated by considering the
effects of placing a stack of $N$ coincident D3-branes in flat ten
dimensional space.  The D3-branes are 3+1 dimensional surfaces on
which the open strings of type IIB string theory may end.  On the one hand,
a low-energy description of a stack of $N$ D3-branes is ${\mathcal
N}=4$ supersymmetric Yang-Mills theory, loosely motivated by the fact
that the $N^2$ open strings which interconnect the D3-branes can be
reinterpreted as gluons.  On the other hand, the D3-branes are massive
objects and in the large $N$ limit substantially warp spacetime.
``Close'' to the D3-branes, the geometry approaches $AdS_5 \times
S^5$.  

Given this correspondence, for which there is now an enormous amount of
evidence but no proof, there exists a dictionary mapping stringy
quantities to gauge theory quantities.  The Yang-Mills coupling
constant $g_{YM}^2$ maps to $4 \pi g_s$ where $g_s$ is the string
coupling constant which measures the probability for strings to split.
The 't Hooft coupling $\lambda \equiv g_{YM}^2 N$ maps to $R^4/
\alpha'^2$ where $R$ is the radius of curvature of both $AdS_5$ and
$S^5$ and $1 / 2 \pi \alpha'$ is the string tension.  One usefulness
of the duality lies in the scaling limit $N \to \infty $ while keeping
$\lambda$ large and fixed.  By taking $N \to \infty$, the string
splitting amplitude is suppressed and strings may be treated
classically.  Taking $\lambda$ large corresponds to keeping the
curvature scale of the geometry very large compared to the string
scale, in which limit low-energy processes in string theory are well
approximated by supergravity.  Thus strongly interacting physics on
the gauge theory side gets mapped to classical general relativity.

In this paper, we will be interested in very high energy scattering
processes for which supergravity is not enough, although it provides a
useful point of departure.  While the original gauge/string dualities
have only CFT duals, certain generalizations
exhibit renormalization group flow and confinement.  The hard-wall
model we consider below is a trivial generalization of the
$AdS_5/{\rm CFT}_4$ correspondence where a confinement scale is introduced
by hand by removing a portion of the interior of $AdS_5$.  We will
consider high energy $2 \to 3$ scattering processes in this cut-off
geometry.

Regge phenomena in gauge theories well-described by the gauge/string
correspondence were studied by a number of authors.  The methods used
in this paper appeared first in the small-$x$ region of deep inelastic
scattering \cite{deepinelastic} and were fleshed out much more fully
in \cite{Brower:2006ea}.  There, following ideas of ref.~\cite{deepinelastic},
the Regge limit of $2 \to 2$ scattering was studied.  In a regime
where $N$ is taken very large and $\lambda$ and $s$ are taken very
large in a correlated way, with $t$ fixed, the scattering amplitudes
are dominated by single-Pomeron exchange.  The Pomeron is a coherent
color-singlet object built from gluons whose properties are universal;
it is the object which is exchanged by any pair of hadrons that
scatter at high energy and large impact parameter.  In the dual string
theory, the Pomeron is not the graviton but the graviton's Regge
trajectory.

While string theory in flat space in the Regge limit exhibits Regge
scaling $s^{\alpha(t)}$, in the gauge/string duality context this
flat-space string amplitude must be corrected by the warping of the
metric and finally convolved with supergravity wave functions
representing the scattered hadrons.  Moreover, as was argued in
ref.~\cite{Brower:2006ea}, when $\ln (s/\Lambda^2)$ ($\Lambda$ of order
the confinement scale) becomes large compared to $\sqrt{\lambda}$, the
local form of the string amplitude becomes inappropriate, and the
local Mandelstam variable $t$ must be promoted to a differential
operator that acts on these supergravity wave functions.  This in turn
leads to diffusive behavior of the scattering strings.  This diffusion
is the strong-coupling analogue of the diffusive behavior seen at
small $\lambda$ in the calculations of BFKL \cite{BFKL}.  In
gauge/string duality, the radial direction of $AdS_5$ has a dual
interpretation as an energy scale in the field theory, and the
diffusion happens in this radial direction as well as in the
directions tranverse to the scattering direction.  In BFKL, the role
of the radial dimension is played by a transverse momentum variable
circulating in the ladder graphs of the Regge resummation.

In this paper, we apply the methods of ref.~\cite{Brower:2006ea} to a
more elaborate problem.  We study $2 \to 3$ scattering, in a regime
where two Pomerons are emitted by the initial-state particles and fuse
to make a third particle.  Here we will consider this particle to be a
glueball: a normalizable state in the $AdS_5$ space.  Experimentally
one might also be interested in quarkonium states, or a Higgs boson,
but we will not consider these in this paper.

We begin in section \ref{sec:kinematics} by
reviewing the kinematics for these $2 \to 3$ processes.  Instead of
two independent Mandelstam variables $s$ and $t$, five-point
amplitudes involve the variables $s$, $s_1$, $s_2$, $t_1$, and $t_2$.
The double diffractive limit, or double Regge limit, we consider consists of
taking $s$ and $s_i$ very large compared to $\Lambda^2$ while
keeping the $t_i$ small.  In double diffractive scattering in the
center-of-momentum frame, the hadrons scatter by very small angles.  Since
we will explore this $2 \to 3$ process in the hard-wall model, 
we review details of this model in section \ref{sec:hardwall}.
We also present the glueball and hadron wave functions that will be
used in the calculations.

The next step is to construct the five point, flat space, string
theory amplitude which we do in appendix \ref{sec:fivepointcalc}.
While the four point amplitude, or Virasoro-Shapiro amplitude, is well
known and relatively simple, the corresponding closed string five
point amplitude is vastly more complicated and less well known.  In
general, using the techniques of \cite{KLT}, the five point amplitude
can be expressed as a quadratic polynomial in generalized
hypergeometric functions ${}_3 F_2$.  Fortunately, we only need this
amplitude in the double diffractive limit, and our discussion is taken
in large part from \cite{Lipatov:1987nn}.

In section \ref{sec:Smatrix}, given the flat space amplitude in the
double diffractive limit, we use the procedure outlined in
\cite{Brower:2006ea} to convert this flat space amplitude into a
curved space amplitude.  The amplitude may be expressed as an integral
of the glueball wave function over a source function ${\cal R}$ which
is a property of the two fusing Pomerons.  Many properties of
this $2 \to 3$ process can be evaluated independently of the produced
fifth particle by studying ${\cal R}$.

In section \ref{sec:limits} we pause to explain the constraints placed
on the various parameters of the scattering process by the kinematics.

We evaluate this double diffractive scattering amplitude in various
regimes of $t_i$ and $s_i$ in section \ref{sec:variousregimes}.
We begin by considering $t_i=0$ for large values of $\ln (s/\Lambda^2)$ and
$\ln (s_i/\Lambda^2)$.  In this case, we find that the scattering amplitude, an
even function of rapidity, $y$, is a function only of the combination
$y / \sqrt{\lambda}$.  In other words, at large $\lambda$, the
amplitude is nearly independent of $y$.  This is consistent with the
corresponding dual gravity amplitude.

Next, we consider arbitrary $t_i$ and $\ln (s/\Lambda^2)$, $\ln (s_i
/\Lambda^2 ) \gg 1$.  For large $t_i$ we find power-law behavior
fall-off with $t_i$, without any sign of the exponential Regge peak in
the forward region.  The Regge peak is only observed for smaller
values of $t_i$ and $\ln (s_i/\Lambda^2)$, as a transient rather than an
asymptotic phenomenon.  This is roughly consistent with the hard
scattering results of \cite{HS}.

Finally, we present a concluding discussion in section \ref{sec:discussion}.

\section{Double diffractive kinematics of $2 \goesto 3$}
\label{sec:kinematics}

In this paper, we calculate $2 \goesto 3$ scattering amplitudes 
in the double diffractive limit. We have in mind a process similar to 
$pp \goesto pp+ {\rm glueball}$. The incoming hadrons
are deflected by very small angles from their original trajectories 
and produce a glueball via double Pomeron exchange.  

\begin{FIGURE}[h]{
\centerline{\psfig{figure=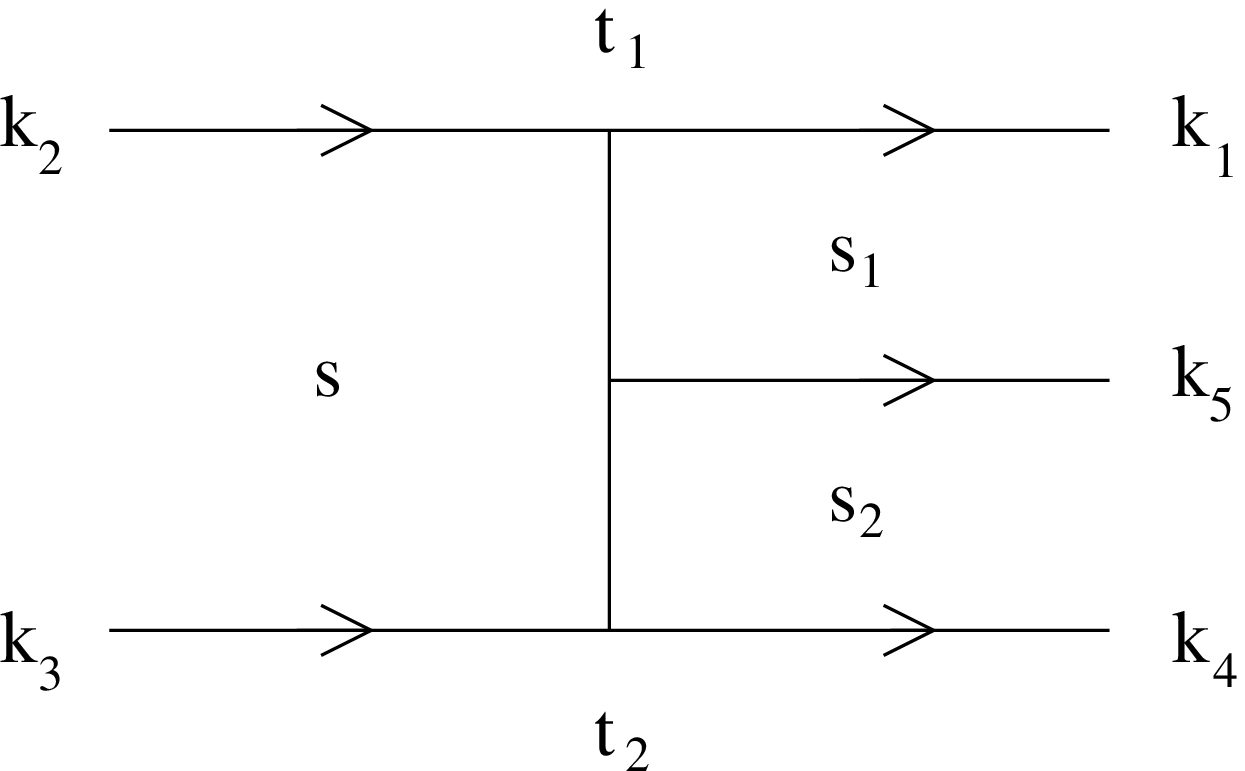,width=2.5in}}
\vspace*{-20pt}
\caption{Momentum flow diagram for double diffractive scattering. 
The hadrons have momenta $k_1, \ldots, k_4$ and the glueball has
momentum $k_5$. Pomerons are represented by the internal lines 
in the $t_1$ and $t_2$ channels.}
\label{fig:2to3}
}
\end{FIGURE}

Working in $-+++$ signature, we will use generalized Mandelstam variables
\begin{eqnarray}
s &=& -(k_2 + k_3)^2, \\
s_1 &=& -(k_5 + k_1)^2, \\
s_2 &=& -(k_4+ k_5)^2, \\
t_1 &=& -(k_1 - k_2)^2, \\
t_2 &=& -(k_3 - k_4)^2,
\end{eqnarray}
and the five masses $m_i^2 = -k_i^2$. Our conventions are shown in 
Figure \ref{fig:2to3}. We will 
assume that the energies are large enough to neglect the external hadron
masses. The glueball has mass $m_5 \equiv m$.

The double diffractive limit we consider is the double Regge limit 
\cite{Brower:1974yv} defined as
\begin{equation}
\label{ddlimit}
s, s_1, s_2 \goesto \infty ~~ {\rm with} ~~ s \gg s_1, s_2 ~~ 
{\rm and} ~~ t_1, t_2, {s_1s_2\over s} ~~ {\rm fixed}.
\end{equation}
In this limit, and going to the center-of-momentum frame, it is 
helpful to parametrize the momenta in the following way: 
\begin{eqnarray}
& k_2 = (E, 0, 0, E), ~~~~~~ k_3 = (E, 0, 0, -E), \nonumber \\
& k_1 = \xi k_2 + \chi k_3 + k_{1\perp}, ~~~~~~ 
k_4 = \xibar k_3 + \chibar k_2 + k_{4\perp}, \nonumber \\
& k_5 = (1 - \xi - \chibar) k_2 + (1 - \chi - \xibar) k_3 
- (k_{1 \perp} + k_{4 \perp}).
\end{eqnarray}
The momenta $k_{1 \perp}$ and $k_{4 \perp}$ are defined to
be orthogonal to $k_2$ and $k_3$.
The eight scalar parameters introduced above must be expressable 
in terms of Lorentz invariants and $m$. Putting momenta on-shell 
and solving algebraically gives
\begin{eqnarray}
& E = \sqrt{s}/2, ~~~~~~ \chi = -t_1/s, ~~~~~~ \chibar = -t_2/s, \nonumber\\
& \xi = 1 + (t_1 - s_2)/s, ~~~~~~ \xibar = 1 + (t_2 - s_1)/s, \nonumber\\
& k_{1\perp}^2 = -t_1 - t_1(t_1 - s_2)/s, ~~~~~~ 
k_{4\perp}^2 = -t_2 - t_2(t_2 - s_1)/s, \nonumber\\
& 2k_{1\perp}\cdot k_{4\perp} = \frac{s_1 s_2}{s}-m^2 
+ \frac{2t_1 t_2}{s} + t_2\(1 - \frac{s_2}{s} \) +t_1 \( 1 - \frac{s_1}{s}\).
\end{eqnarray}
The conditions on the Mandelstam variables in the double diffractive
limit translate into the relations $\xi, \xibar \approx 1$ 
and $\chi, \chibar \ll 1$.  These relations in turn can be understood
physically as the fact that the hadrons are deflected
by very small angles in this limit.

It is also convenient to define 
\begin{equation}
m_{\perp}^2 \equiv m^2 + (k_{1\perp} + k_{4\perp})^2.
\label{mperpdef}
\end{equation}
This is the effective mass of the glueball, as viewed in the $x^+$--$x^-$ 
light-cone plane.
In terms of Mandelstam variables, 
\begin{equation}
m_{\perp}^2 = \frac{s_1 s_2}{s} \( 1 - \frac{ t_1 - t_2}{s_2} \) 
\(1 + \frac{t_1 -t_2}{s_1} \) \approx \frac{s_1 s_2}{s}
\end{equation}
in the double diffractive limit.  

The rapidity $y$ of the glueball is defined by $\tanh y = k_{5z}/E_5$. 
In terms of the Mandelstam variables, the rapidity can be
expressed as
\begin{equation}
y = \ha \ln\({E_5 + k_{5z}\over E_5 - k_{5z}}\) = 
\ha \ln \left( \frac{s_2+t_2-t_1}{s_1+t_1-t_2} \right) 
\approx \frac{1}{2} \ln \left( \frac{s_2}{s_1} \right) \ .
\label{rapiditydef}
\end{equation}
In the center-of-momentum frame, for a massless particle, the 
rapidity is related to the polar angle (relative to the beam axis)
via $e^y = \cot (\theta/2)$:
large $|y|$ corresponds to $\theta \approx 0$ or $\pi$.

\section{Hard-wall model}
\label{sec:hardwall}
The hadronic wave functions that are needed in our computations depend
on how confinement is incorporated into the gauge/string framework. In
this paper, we choose to use the hard-wall model.\footnote{The
hard-wall model is not a fully consistent SUGRA theory as it does not
satisfy the supergravity equations of motion. However, past
calculations have shown that the model is useful for capturing
model-independent behavior of confining gauge theories
\cite{HS,deepinelastic,rhouniv,erlichetal,
daroldpomarol,Brower:2006ea}.}  The model imposes a sharp cutoff on
the $AdS_5$ radial coordinate $z$ at some scale $z_0$,
\begin{equation}
\label{metric}
ds^2 = {R^2\over z^2}(\eta_{\mu\nu} dx^\mu dx^\nu + dz^2) + ds_W^2, 
~~~~ 0<z < z_0.
\end{equation}
The line element $ds_W^2$ gives the metric on the five-dimensional
transverse space, which in simple examples is just $S^5$.
The cutoff establishes a mass gap $\Lambda = 1/z_0$. To a 
four-dimensional observer, $z \to 0$ is the UV, while $z \to z_0$ 
is the IR.  

Scalar hadron wave functions are solutions to the ten-dimensional 
Klein-Gordon equation using the metric (\ref{metric}). By separation 
of variables the solutions can be written as 
$e^{ik\cdot x} \phi(z) f(\th)$, where $k$ is the gauge theory 
four-momentum. For simplicity, we take $f$ to be a constant mode 
on the compact space $W$. The function $f$ plays no role in our 
calculations.  
When we introduce diffusion kernels, it will be convenient to use 
the dimensionless coordinate $u = \ln(z_0/z)$, $0 \leq u < \infty$. 
To a four-dimensional observer, $u \goesto 0$ is the IR and 
$u\goesto\infty$ is the UV. 

The radial-``Kaluza-Klein'' normalizable mode corresponding to a hadron 
is given by \cite{Hong:2004sa}
\begin{equation}
\label{wavefnu}
\phi(u) 
= {\sqrt{2}\over\La\sqrt{{\rm Vol}_W}R^{3/2}}
e^{-2u}{J_{\D-2}([m/\La]e^{-u})\over J_{\D-2}(m/\La)},
\end{equation}
where ${\rm Vol}_W$ is the volume of the compact space $W$,
$\D$ is the conformal dimension of the dual gauge theory
operator, and $m$ is the four-dimensional mass given by 
$m / \Lambda = \zeta_{\D-3;n}$. $\zeta_{k;n}$ denotes the $n$th zero 
of the Bessel function $J_k(x)$.\footnote{We use the generalized 
Neumann boundary conditions
$\partial_u \left(e^{-(\Delta-4)u}\phi(u)\right) = 0$
at $u=0$ employed in \cite{Hong:2004sa}, which imply
$\phi'(0) = (4-\Delta) \phi(0)$.  
These reduce to ordinary Neumann boundary conditions
only in the case $\Delta=4$.
The normalization is fixed by the condition ${\rm Vol}_W R^3
\Lambda^2 \int_0^\infty du\, e^{2u} \phi(u)^2 = 1$. See the appendix
of ref. \cite{HS} for a derivation of this normalization condition.
}

In Figure \ref{fig:2to3}, the hadron with momentum
$k_5$ represents the glueball. In non-Abelian gauge theory, 
a scalar glueball can be created by the operator 
${\cal O} = \Tr F_{\mu\nu}F^{\mu\nu}$.
According to the $AdS$/CFT correspondence, this operator is dual to a massless
closed string dilaton state propagating in the ten-dimensional bulk
whose $AdS$ radial profile is given by the normalizable mode (\ref{wavefnu}) 
with $\D = 4$.  

The external hadrons with momenta $k_1, \ldots, k_4$ in 
Figure \ref{fig:2to3} are all scalar normalizable modes $\phi_0$
of the form (\ref{wavefnu}) with $\D = \D_0$ and mass $m_0$.  
Although they are not baryons, and are scalars rather than
fermions, they are reasonable surrogates
for protons, as their profile in the radial $AdS$ direction is 
appropriate for an object built out of a small number
of valence partons (and created by an operator of small
twist \cite{HS}).

\section{S-matrix}
\label{sec:Smatrix}
We want to compute the S-matrix, ${\cal S}$, for hadron scattering in
four dimensions. In string theory, at leading order in $1/N$ and
$1/\sqrt{\lambda}$, this is given roughly by a path integral over a
spherical worldsheet embedded in the cut-off $AdS_5 \times W$ space,
with appropriate vertex operators representing the external hadron
states. For scattering processes in $d$ flat dimensions, we also
define the amplitude ${\cal T}_d$ by 
\begin{equation}
\label{amplitude}
{\cal S} = i(2\pi)^d \d^d(\sum k_i) {\cal T}_d.
\end{equation}
Using a prescription given in \cite{Brower:2006ea} we determine
${\cal S}$ by integrating ${\cal T}_{10}$ (the scattering amplitude
for closed strings in ten-dimensional Minkowski space) with the
product of hadron wave functions over all coordinates in the cut-off
$AdS_5 \times W$,
\begin{equation}
\label{S}
{\cal S} = \int_{AdS_5} d^4x\, dz \int_W d^5\th \sqrt{-G} ~ 
{\cal T}_{10}(\widetilde{k}_1, \ldots, \widetilde{k}_5)
\prod_{i = 1}^5 e^{ik_i\cdot x}\phi_i(z).
\end{equation}
The factor $\sqrt{-G}$ is the square root of the determinant of the
metric on $AdS_5 \times W$.
It is important to distinguish the two sets of momenta that arise 
in (\ref{S}). The $k_i$ are the four-momenta of scalar hadrons 
in the dual gauge theory defined on the boundary of $AdS_5$, 
while the $\widetilde{k}_i$ can be interpreted as (center-of-mass) 
ten-momenta of closed strings propagating in the $AdS_5 \times W$
background at a given value of $z$. 
The relationship between $k_i$ and $\widetilde{k}_i$
will be discussed later in this section.

Our first task is to compute the tree-level string 
amplitude ${\cal T}_{10}$ for five external states.  Rather 
than use the closed superstring, we calculate in appendix 
\ref{sec:fivepointcalc} an equivalent amplitude 
${\cal T}_{26}$ for the closed bosonic string.  In the double 
Regge limit, where exchange of the leading Regge
trajectory in the $\widetilde{t}_1$ 
and $\widetilde{t}_2$ channels dominates, we expect bosonic and 
superstring amplitudes to be essentially identical, up to overall constants
and subleading effects
which are not important for us here.  
Moreover, in this limit the amplitude
will not depend sensitively on the precise vertex operators used
to represent our hadrons on the external legs.  In particular,
only their energy-momentum tensor is important, as a source for
the emitted Pomeron.  Thus
there is no cost in using tachyons as the initial state vertex
operators in the $2\to3$ scattering process.  We must be more precise
about the fifth particle (the ``glueball'') whose coupling to 
gravitons, and therefore to two Pomerons, contributes nontrivial
kinematic factors to the amplitude.  We use the dilaton as the
particle produced by the colliding Pomerons; this
corresponds to producing a scalar glueball in the gauge theory.


Thus, we calculate the bosonic closed string amplitude ${\cal T}_{26}$ 
for four tachyons and one dilaton in the double Regge limit, 
with the dilaton singled out as the produced particle.
In appendix \ref{sec:fivepointcalc}, it is shown that ${\cal T}_{26}$
can be written as a power series in the
dimensionless parameter $\a'\widetilde{m}^2_\perp$.  Here
$\widetilde{m}^2_\perp$, the ten-dimensional generalization of
$m_\perp^2$, is approximately a ratio of ten-dimensional Mandelstam
invariants, $\widetilde s_1\widetilde s_2/\widetilde s$, in the double
Regge limit.  As we will see in section \ref{sec:limits}, 
the physics forces us into a regime where
this parameter is much less than 1. 
The leading term in the series for ${\cal T}_{26}$ is given in 
eq.~(\ref{T26result}).
We modify the normalization of (\ref{T26result}) so that it may
be interpreted as a ten-dimensional scattering 
amplitude, taking $g_c \sim \alpha'^2 g_s$.
To leading order in 
$\a'\widetilde{m}^2_\perp$ and up to a numerical factor,
\begin{eqnarray}
\label{T}
{\cal T}_{10} & \sim & 
\a'^5 g_s^3 \\
& & \times \left[~\, 
\(e^{-i\pi/2}{\a'\widetilde{s}\over4}\)^{2+\a'\widetilde{t}_1/2} 
\(e^{-i\pi/2}{\a'\widetilde{s}_2\over4}\)^{
\a'(\widetilde{t}_2-\widetilde{t}_1)/2}
\Pi(\a'\widetilde{t}_1/4,\a'(\widetilde{t}_2-\widetilde{t}_1)/4)\right. 
\nonumber\\
& & \left. ~~\; + 
\(e^{-i\pi/2}{\a'\widetilde{s}\over4}\)^{2+\a'\widetilde{t}_2/2} 
\(e^{-i\pi/2}{\a'\widetilde{s}_1\over4}\)^{
\a'(\widetilde{t}_1-\widetilde{t}_2)/2}
\Pi(\a'\widetilde{t}_2/4,\a'(\widetilde{t}_1-\widetilde{t}_2)/4) \right]
\nonumber
\end{eqnarray}
where
\begin{equation}
\label{propagator}
\Pi(x,\d) = (x+\d) {\G(-1-x)\over \G(2+x)}{\G(-\d)\over \G(1+\d)}.
\end{equation}

It was shown in \cite{Brower:2006ea}, following \cite{HS,deepinelastic}, 
that a string amplitude in flat
spacetime can be carried over in some circumstances
to an amplitude in a weakly-curved 
spacetime. To do this one must relate the ten-dimensional 
Lorentz invariants in the string 
theory: $\widetilde{s}, \widetilde{s}_1, \widetilde{s}_2, 
\widetilde{t}_1,\widetilde{t}_2$, to the four-dimensional invariants 
in the gauge theory: $s, s_1, s_2, t_1, t_2$. We define an effective 
string length-squared as $\a'$ divided by the $AdS$ warp 
factor,\footnote{This combination arises naturally in the worldsheet 
path integral. Separating bosonic fields into constant zero modes 
plus stringy fluctuations, e.g., $Z(\s^1,\s^2) = z + Z'(\s^1,\s^2)$, 
introduces an overall factor of $R^2/\a'z^2$ in front of the 
worldsheet action.}
\begin{equation}
\a'_{\rm eff}(z) \equiv \a'z^2/R^2 = z^2/\sqrt{\la}.
\end{equation}
Each kinematic quantity $\tilde s, \tilde s_i, \tilde t_i$ in the string
theory should be understood as standing in for a differential 
operator, a Laplacian acting on the appropriate states.
The eigenvalue of this operator in flat space would be the
usual kinematic quantity in the string theory, but in curved
space the Laplacian is not trivial.  For example, the quantity
$\a'\tilde t_1$ is a differential operator 
\begin{equation}
\label{t1tilde}
\a'\widetilde{t}_1  \equiv  \a'\nabla^2_{P_1} \equiv {1\over\sqrt\lambda}
\left[z {\p \over \p z} z {\p \over \p z} +
t_1 z^2 - 4\right] + O(1/\la)  \ .
\end{equation}
This operator, the covariant spin-2 Laplacian appropriate for the
Pomeron being exchanged in the $t_1$ channel
\cite{deepinelastic,Brower:2006ea}, acts on states 1 and 2; more
precisely, it acts on the product of their wave functions.  Similar
expressions apply for $\widetilde t_2$, $\widetilde s$ and $\widetilde
s_i$, but $s$, $s_1$ and $s_2$ are so large compared to the remainder
of the differential operator that we may approximate
\begin{equation}
\label{s1tilde}
\a'\widetilde{s}_1  \approx
{1\over\sqrt\lambda} s_1 z^2 \equiv \a'_{\rm eff}(z) s_1 \ .
\end{equation}
and similarly for $s_2$ and $s$.
In summary, we take
\begin{eqnarray}
\label{holo}
\a'\widetilde{s} & \equiv & \a'_{\rm eff}(z)s, \nonumber\\
\a'\widetilde{s}_i & \equiv & \a'_{\rm eff}(z)s_i, \nonumber\\
\a'\widetilde{t}_i & \equiv & \a'\nabla^2_{P_i} .
\end{eqnarray}
It is essential to retain the full differential operator in the last
expresssion because $\a'\widetilde{t}_i$ appears in the exponent of
$\a'\widetilde{s}$, and $s$ is taken to be exponentially large in
$\sqrt{\la}$.

We can study the properties of $\a'\nabla^2_{P_i}$ by using
coordinates in which the operator becomes the Hamiltonian for a
quantum mechanics problem, in which an analogue particle moves
in a potential.  Let $u = \ln(z_0/z)$ and define
\begin{equation}
\label{diffusiveH}
H_i = -\sqrt{\la} \a' \nabla^2_{P_i} = -{\p^2\over\p u^2} + V_i(u),
\end{equation}
where, for the $AdS$ metric with $z=z_0 e^{-u}$, the
effective potential takes the form
\begin{equation}\label{V}
V_i(u) = 4 - z_0^2 t_i e^{-2u} \ . 
\end{equation} 
For large negative $t_i$ the potential
grows exponentially  as $u$ decreases.  In a confining theory,
where confinement physics modifies the infrared (small $u$, large $z$)
region, this exponential potential provides an infrared cutoff
that screens the details of the physics of confinement.  As $t_i\to0$
this screen is removed and the precise nature of confinement
comes into play.  

Within the simplistic but useful hard-wall model,
the potential takes the form (\ref{V}) all the way to $u=0$, where
the space ends, with an appropriate boundary condition.
In this case the quantum mechanics problem is completely solved for
any $t_i$.  
A complete set of 
eigenfunctions $\psi_\nu(t_i, u)$ with eigenvalues $E = 4 + \nu^2$ is
given in appendix \ref{sec:eigfuncs}.  

In many circumstances it is useful to write the amplitude explicitly
in terms of these eigenfunctions.  For an arbitrary 
functional $F$ we use completeness of the eigenfunctions to write
\begin{eqnarray}
\label{inttransform}
F[\a'\nabla^2_{P_1}] \phi_1(z)\phi_2(z) & = & 
F[-H_1/\sqrt{\la}] \phi_1(u)\phi_2(u) \\
& = & \int_0^\infty du' \phi_1(u')\phi_2(u') \int_0^\infty d\nu 
F\(-{4+\nu^2\over\sqrt{\la}}\) \psi_\nu(t_1,u)\psi^*_\nu(t_1,u'). \nonumber
\end{eqnarray}
$\nabla^2_{P_1}$ is defined to act only on the product
$\phi_1\phi_2$. Likewise, $\nabla^2_{P_2}$ acts only on $\phi_3\phi_4$.

Using (\ref{inttransform}) it is straightforward to express (\ref{S}) 
explicitly in the hard-wall model as
\begin{eqnarray}
\label{Smatrix}
{\cal S} & = &  (2\pi)^4\d^4\(\sum k_i\)
{\cal N}\a'^5 g_s^3\ {\rm Vol}_W R^5\nonumber \\
& & \times
\int_0^\infty du\,{e^{4u}\over z_0^4} \phi_5(u) 
\int_0^\infty du' \phi_1(u')\phi_2(u')
\int_0^\infty du'' \phi_3(u'')\phi_4(u'')
D(u,u',u''),
\end{eqnarray}
where 
\begin{eqnarray}
D(u,u',u'') 
& = & 
\(e^{-i\pi/2}{\a'_{\rm eff}(u)s\over 4}\)^{2-2/\sqrt{\la}} D_0(u,u',u'') 
\label{D}\\
D_0(u,u',u'') 
& = & 
\int_0^\infty d\nu\, \psi_\nu(t_1,u)\psi_\nu^*(t_1,u')
\int_0^\infty d\w\, \psi_\w(t_2,u)\psi_\w^*(t_2,u'') \nonumber \\
& & 
\times \[\,~~ e^{-\t \nu^2} e^{\t_2(\nu^2-\w^2)} 
\Pi\(-{4+\nu^2\over4\sqrt{\la}},{\nu^2-\w^2\over4\sqrt{\la}}\)
\right.\nonumber\\
& & 
\left.~~~ + e^{-\t \w^2} e^{\t_1(\w^2-\nu^2)} 
\Pi\(-{4+\w^2 \over4\sqrt{\la}},{\w^2-\nu^2\over4\sqrt{\la}}\)~\] \ .
\label{D0a}
\end{eqnarray}
Here ${\cal N}$ is a normalization factor which, among
other things, will
correct for the fact that we have used bosonic strings in place 
of superstrings to formulate the S-matrix. 
The variables appearing in the exponentials of 
(\ref{D0a}) are
\begin{equation}
\label{diffusiontimes}
\t = {1\over2\sqrt{\la}}\[\ln(\a'_{\rm eff}(z)s/4) - i\pi/2\], ~~~~
\t_i = {1\over2\sqrt{\la}}\[\ln(\a'_{\rm eff}(z)s_i/4) - i\pi/2\].
\end{equation}
These are all functions of $z$, but vary slowly with $z$.  They 
are analogous to diffusion times \cite{Brower:2006ea}.  
Since the integrations in
(\ref{Smatrix}) are over regions in which $z$ is not exponentially
small in $\sqrt{\la}$, this variation is subleading,
as long as we choose to evaluate $\a'_{\rm eff}(z)$ at a sensible
place, where no large logarithms arise.  It is natural to evaluate
(\ref{diffusiontimes}) where the integrand of (\ref{Smatrix}) is
peaked; we will refer to this value of $z$ as $z_{\rm scatt}$.
In the double Regge limit, the $s$-type Mandelstam variables are 
exponentially large in $\sqrt{\la}$, so the diffusion times have 
positive real parts.

The interpretation of ${\cal S}$ given by (\ref{Smatrix}) is 
straightforward (reading right to left).  We convolve pairs of 
hadron wave functions in the $t_1$ and $t_2$ channels with the 
diffusion kernel, and take an overlap of the result with the 
glueball wave function.
Indeed, it will be useful
to define 
the quantity
\begin{equation}
\label{Pomeron}
{\cal R}(u) \equiv 
{e^{4u}\over z_0^4}
\int_0^\infty du'\,
\phi_1(u')\phi_2(u')
\int_0^\infty du''\,
\phi_3(u'')\phi_4(u'')
D(u,u',u'') \ .
\end{equation}
In terms of ${\cal R}$, the S-matrix is simply
\begin{equation}
\label{SRrule}
{\cal S} = 
(2\pi)^4 \d^4\(\sum k_i\)
{\cal N}\a'^5 g_s^3 {\rm Vol}_W R^5
\int_0^\infty du\, \phi_5(u) {\cal R}(u).
\end{equation}
Thus the function ${\cal R}$, which depends on the specific external
hadrons chosen and on the kinematics, but not on the produced glueball, 
may be interpreted as a ``source'' for the glueball state.  We will
refer to it here as the ``double Pomeron source function.''

In (\ref{D0a}) we may assume that the size of $\t$ is sufficient to
guarantee that only the eigenvalues $\nu,\w \ll \la^{1/4}$ are 
important for the evaluation of the integral. Thus, the arguments of $\Pi$ 
are small and we can expand $\Pi$,
\begin{equation}
\lim_{x,\d\goesto 0} \Pi(x,\d) \sim -{1\over x} - {1\over \d}.
\label{proplimit}
\end{equation}
Rearranging yields
\begin{eqnarray}
\label{D0b}
D_0(u,u',u'') 
& \approx & 
-4\sqrt{\la}\int d\nu\, \psi_\nu(t_1,u)\psi_\nu^*(t_1,u')
\int d\w\, \psi_\w(t_2,u)\psi_\w^*(t_2,u'') \nonumber \\
& & 
\times\, e^{-\t_1\nu^2 -\t_2\w^2}\left[
{e^{\t_\perp\nu^2} - e^{\t_\perp\w^2}\over \nu^2-\w^2} - 
{e^{\t_\perp\nu^2}\over \nu^2+4} - 
{e^{\t_\perp\w^2}\over \w^2+4}\right],
\end{eqnarray}
where 
\begin{equation}
\label{tauperp}
\t_\perp \equiv -\t + \t_1 + \t_2 \approx 
{1\over2\sqrt{\la}}\[\ln(\a'_{\rm eff}(z_{\rm scatt})m^2_\perp/4) - i\pi/2\].
\end{equation}

\section{Discussion of Parameters and Limits}
\label{sec:limits}
Our version of the double diffractive process $pp \goesto pp + {\rm glueball}$ 
is fully described by the choice of six parameters
$N, \la, \D_0, m_0, \D, m$, and the five kinematic variables
$s, s_1, s_2, t_1, t_2$.

In this work the parameters are constrained as follows.
The number of colors $N$ must be large to ensure that the S-matrix 
is dominated by the lowest genus worldsheet. The 't Hooft coupling 
$\la$ must be large to be consistent with our calculation of closed
strings propagating in an AdS background with a large radius of 
curvature. For the glueball, $\D = 4$ and $m/\La$ can be
any zero of the Bessel $J_1$ function.

There are also important constraints on the kinematic variables.
In the physical region, the momenta transfer-squared $t_1$ and $t_2$ are 
negative semidefinite in any scattering process.  
For Regge behavior to be relevant, we will need 
$|s|, |s_1|, |s_2|$ to be very large compared to $\Lambda^2$; this requires
the parameters $\t, \t_1, \t_2$ to be large compared to $1/\sqrt\lambda$.

One apparent 
assumption constraining the kinematics arises in the
calculation of the amplitude ${\cal T}_{26}$, where we assumed
$\a'\widetilde{m}^2_\perp \ll 1$ in order to keep just the first
term in a power series solution.  According to (\ref{holo}), in a
warped metric this becomes the condition
\begin{equation}\label{condition1}
\a'_{\rm eff}(z)m^2_\perp = 
{(z/z_0)^2\over\sqrt{\la}}{m_\perp^2\over\La^2} \ll 1.
\end{equation}
This would obviously be satisfied for all $z$ if we 
were to impose $m_\perp^2/\La^2 
\ll \sqrt{\la}$, or equivalently, ${s_1 s_2/ s} \ll \La^2 \sqrt{\la}$.
However, this condition would be much stronger than necessary.
In fact, the physical process itself assures that 
$\alpha'_{\rm eff}(z) m_\perp^2$ is never larger than one, 
without additional assumptions.  

The logic is the following.  If we produce
a glueball with mass of order $\Lambda$, then $m_\perp^2$ itself can
only be large if either $-t_1\gg \Lambda^2$ or $-t_2\gg\Lambda^2$.
This follows from the definition (\ref{mperpdef}), which, along with
the relations $t_1\approx -k_{1\perp}^2$, $t_2\approx -k_{4\perp}^2$,
assures that
\begin{equation}\label{mperpinequality}
m^2 < m_\perp^2 < m^2 + 4 \, \mbox{max}(|t_1|, |t_2|) \ .
\end{equation}
Without loss of generality, let us assume that $|t_1| > |t_2|$.
But if $-t_1\gg\Lambda^2$, then it serves as an infrared cutoff
on the dynamics, causing $\mathcal{R}(z)$ to have support only
at regions of $z$ which are small compared to $1/\sqrt{-t_1}$.  
In our computation this arises from the potential (\ref{V}),
which develops a $-t_1 e^{-2u}\propto -t_1 z^2$ barrier at 
large $z$, forcing the physics to smaller $z$.  In particular, all
the eigenfunctions $\psi_\nu$ from which $\mathcal{R}(z)$
is built have exponentially falling tails for
$z > 1/\sqrt{|t_1|}$.
Therefore, if we take $m_\perp^2$ large by making $|t_1|$
large, we find 
\begin{equation}
\label{tildemperpconstraint}
\alpha' \widetilde{m}_\perp^2 = \frac{z^2}{\sqrt{\lambda}} m_\perp^2 
< \frac{1}{|t_1|} \frac{m_\perp^2}{\sqrt{\lambda}}  
< \frac{4}{\sqrt{\lambda}} \ll 1
\end{equation}
where we have used eq.~(\ref{mperpinequality}).
Thus in our calculations we do not need to impose eq.~(\ref{condition1}),
because it follows automatically from the dynamics.

One corollary of this discussion is that $\t_\perp \ll 1$.  Clearly,
from the definition of $z_{\rm scatt}$ below (\ref{diffusiontimes}), 
it cannot be that
$\mathcal{R}(z_{\rm scatt})$ is exponentially small, except possibly in
regions that make exponentially small contributions to amplitudes.
Therefore it follows that the above constraint on
$\alpha'\widetilde{m}_\perp^2$ applies.  From its definition (\ref{tauperp}),
combined with (\ref{tildemperpconstraint}), $\t_\perp$ is therefore
always of order $1/\sqrt{\lambda} \ll 1$ without further assumption.

\section{Calculation in various kinematic regimes}
\label{sec:variousregimes}
Our goal in this section is to compute the double Pomeron source
function ${\cal R}(u)$. We do this first for the limiting case of
vanishing momentum transfers, where we will observe a minimal rapidity
dependence of the amplitude, and infrared insensitivity for
sufficiently large $s, s_1, s_2$.  We then consider the case of
negative momentum transfers, first observing the absence of classic
Regge behavior at large $\t_i$, and then identifying it at smaller
$\t_i$ as a transient phenomenon.  We also find power-law fall-off at
large $\tau_i$ and large $t_i$, analogous to what was seen
in \cite{HS}.

\subsection{Zero $t_1, t_2$ and large $\tau_i$}
\label{sec:zerot}
We begin by calculating the double Pomeron source function for
$t_1=t_2=0$. 
In this regime no infrared cutoff protects the
scattering process from the details of confinement, and we do not
expect the detailed results from the hard-wall model to apply
generally to all theories.  Nevertheless, we may hope for universal
behavior in some limited settings.  We will find that the source
function is nearly flat in rapidity, for reasons which should apply
to all theories at large 't Hooft coupling.  We will also see
that the source function becomes less sensitive to the infrared as the
center-of-momentum energy increases.  While this would be very interesting
if it were universal, we unfortunately see no reason why this
should be the case.

In order to find ${\cal R}$ we must first calculate the diffusion
kernel according to eq.~(\ref{D0b}). This formula calls for the 
eigenfunctions of the hard-wall model Hamiltonian at $t_1 = t_2 =
0$, which are given explicitly in the appendix by eq.~(\ref{zeroteigfuncs}).
Upon simplifying, one finds that the kernel can be written in terms of
the following integrals:\footnote{Integrals $P$ and $Q$
may be calculated using the method of Fourier transforms combined
with the convolution theorem.}
\begin{eqnarray}
P(u,u',\t) & \equiv & 
\int_0^\infty d\nu\, \psi^*_\nu(u)\psi_\nu(u') e^{-\t\nu^2} \nonumber \\
& = & {e^{-u_-^2/4\t}\over2\sqrt{\pi\t}} + 
{e^{-u_+^2/4\t}\over2\sqrt{\pi\t}}
\[1 - 4\sqrt{\pi\t} f\({u_++4\t\over\sqrt{4\t}}\)\], \label{P}
\end{eqnarray}
and
\begin{eqnarray}
Q(u,u',\t) & \equiv & 
\int_0^\infty d\nu\, \psi^*_\nu(u)\psi_\nu(u') {e^{-\t\nu^2}\over\nu^2+4} 
\nonumber \\
& = & {e^{-u_-^2/4\t}\over8}\[f\({-u_-+4\t\over\sqrt{4\t}}\)+
f\({u_-+4\t\over\sqrt{4\t}}\)\] \nonumber \\
& & - e^{-u_+^2/4\t}\sqrt{\t}\[{1\over\sqrt{\pi}}-
\({u_++4\t\over\sqrt{4\t}}\)f\({u_++4\t\over\sqrt{4\t}}\)
\]. \label{Q}
\end{eqnarray}
Here we have defined $u_\pm = u \pm u'$ and $f(x) = e^{x^2}{\rm erfc}(x)$,
with the convention ${\rm erfc}(x) \equiv 1 - 
(2/\sqrt{\pi}) \int_0^x \exp(-t^2) dt$. We now have 
\begin{equation}\label{D0approx}
{D_0\over4\sqrt{\la}} \approx P(u,u',\t_1)Q(u,u'',\t_2) 
+ Q(u,u',\t_1)P(u,u'',\t_2) + O(\t_\perp). 
\end{equation}
%
%
%
%
%
%
The diffusion kernel is then
\begin{equation}
D(u,u',u'') \approx 4\sqrt{\la}\(e^{-i\pi/2}{\a'_{\rm eff}(u)s\over
4}\)^{2-2/\sqrt{\la}} \[P(u,u',\t_1)Q(u,u'',\t_2) 
+ Q(u,u',\t_1)P(u,u'',\t_2)\].
\end{equation}
If we define
\begin{eqnarray}
\Pbar(u,\t_i) & = & \int_0^\infty du' \phi_0(u')^2 P(u,u',\t_i), \\
\Qbar(u,\t_i) & = & \int_0^\infty du' \phi_0(u')^2 Q(u,u',\t_i),
\end{eqnarray}
then the double Pomeron source function takes the simple form
\begin{equation}
\label{Rpbarqbar}
{\cal R}(u) \approx {4\sqrt{\la}\over z_0^4}
\(e^{-i\pi/2}{z_0^2 s\over4\sqrt{\la}}\)^{2-2/\sqrt{\la}}e^{4u/\sqrt{\la}}
\[\Pbar(u,\t_1)\Qbar(u,\t_2)+\Qbar(u,\t_1)\Pbar(u,\t_2)\].
\end{equation}
By evaluating $\Pbar$ and $\Qbar$ we can obtain an explicit formula
for ${\cal R}$. Since 
$\p_{\t_i} Q - 4Q + P = 0$, we need only determine $\Qbar$, and use 
$\Pbar = 4\Qbar - \p_{\t_i} \Qbar$.  For the hard-wall model the
hadron wave functions are Bessel functions, and we can obtain an exact
expression for $\Qbar$, given in appendix \ref{sec:Qbar}.

It is useful to rewrite the expressions above in
terms not of $\t_1$ and $\t_2$ but $\t_1\pm \t_2$.  
Eq.~(\ref{tauperp}) implies that $\t_1 + \t_2 =
\tau +\t_\perp\approx \tau$, since $\tau_\perp$ is very small.  
In the double diffractive limit, the difference $\t_2-\t_1$ is
proportional to the rapidity $y$ of the glueball, defined in
(\ref{rapiditydef}),
\begin{equation}
y = \sqrt{\la}(\t_2-\t_1) \ ,
\end{equation}
where we used $|t_1|,|t_2|\ll s_1, s_2$.  Since $\t_i$ is positive the
range of allowed rapidity is finite but very large, $-\t\sqrt{\la} < y
< \t\sqrt{\la}$.

By construction ${\cal R}$ is
symmetric in $\t_1$ and $\t_2$, so it must be an even function of 
rapidity.  Meanwhile the rapidity $y$ can only
appear in the ratio $y/\sqrt{\la}$. {\it This means that the shape of 
${\cal R}(u)$ is nearly constant for rapidities $y \ll \sqrt{\la}$.}
This follows essentially from the fact that the gravitational scattering
amplitude to which our calculation reduces in the 
$\lambda\to\infty$ limit is 
rapidity-independent.  

Let us now discuss the main features of the double Pomeron source
function. ${\cal R}(u)$ vanishes
in the UV as $u\goesto \infty$. For large $u$, the Gaussians from $\Qbar$ and
$\Pbar$ (see eqs.~(\ref{Qbarformula}) and (\ref{summand})) dominate giving
\begin{equation}
\label{gaussianfalloff}
{\cal R}(u) \sim \exp \left[
- \frac{u^2}{4} \left( \frac{1}{\tau_1} + \frac{1}{\tau_2} \right) \right]
=
\exp\({-\t u^2\over\t^2 - y^2/\la}\) \ .
\end{equation}

Many aspects of ${\cal R}(u)$ that are true throughout the Regge
regime can be understood qualitatively
by considering the limit of 
large diffusion times $\t, \t_1, \t_2 \gg 1$, and focusing on the
behavior at small
$u \ll \tau_i$.   (The behavior at larger $u$ is typically irrelevant
for our computation, because of the exponential fall-off of the
glueball wave function at large $u$.)
In this limit, we have from eq.~(\ref{remarkable}) that the $u$ 
and $\tau_i$ behavior of ${\cal R}(u)$ is given by
\begin{equation}
\label{Rutdep}
{\cal R}(u) \sim \frac{(1+2u)^2}{(\tau_1 \tau_2)^{3/2}}
 \exp \[-\frac{u^2}{4} \(\frac{1}{\tau_1} + \frac{1}{\tau_2} \)\] 
 \ .
\end{equation}
The $\tau_i^{-3/2}$ behavior was explored in
ref.~\cite{Brower:2006ea}.  At $t_i=0$ it arises from the reflection
off the hard-wall barrier.  For modes with slow variation in $u$,
which dominate at large $\tau_i$, the boundary condition required by
energy-momentum conservation forces the incoming and
outgoing waves to interfere destructively, cancelling the leading
$\tau_i^{-1/2}$ behavior expected in diffusion.  This destructive
interference persists for $t_i \ll - \Lambda^2$, as the exponential
behavior of the effective potential acts as a Dirichlet boundary
condition. (Note however that at extremely large $u\gg \tau_i$ the
inverse square-root behavior is not cancelled.)

As a function of $u$, ${\cal R}$ will have a peak at 
$u_{\rm max} = \frac{1}{4}[\sqrt{1+64\t_1\t_2/\t}-1]$.
When the glueball has rapidity $y=0$, the peak location 
scales as $u_{\rm max} \sim \sqrt{\tau}$, so that the height of the peak
scales as ${\cal R}(u_{\rm max}) \sim \tau^{-2}$. Meanwhile, at the 
hard-wall, ${\cal R}(0) \sim \tau^{-3}$.  The peak-to-wall ratio 
is
\begin{equation}\label{Rmax}
{{\cal R}(u_{\rm max})\over {\cal R}(0)} \approx {4\over e}\tau.
\end{equation}
At non-zero rapidity, keeping the leading correction 
$|y| \ll \tau\sqrt{\la}$, the peak moves to $u_{\rm max} \sim 
\sqrt{\t}[1-(y^2/2\t^2\la)]$ and the ratio (\ref{Rmax})
scales as $\frac{4}{e}\tau[1-(y^2/2\t^2\la)]$.

We learn two important facts.
At fixed central rapidity, and as $\tau$ increases, the
maximum of $\mathcal{R}(u)$ moves away from the confinement region
near $u=0$.  This means that its computation
becomes increasingly insensitive to the details of confinement.  
Furthermore, the ratio of the
maximum of $\mathcal{R}(u)$ to its value in the confining region
becomes large. This means that the confining region plays a smaller
and smaller role in the computation of the source function.  
Unfortunately, we currently
see no argument that the insensitivity of this computation to the
confining regime should apply generally to most or all large-$\lambda$ gauge
theories. Instead, it appears to be a special
feature of the hard-wall model.

In Figure \ref{fig:R_t0} we plot a dimensionless version of the 
double Pomeron source function: 
\begin{equation}\label{dimlessR}
\widehat{\cal R}(u) \equiv {z_0^4 \over 4\sqrt{\la}}
\(e^{-i\pi/2}{z_0^2 s\over 4\sqrt{\la}}\)^{-2+2/\sqrt{\la}}
e^{-4u/\sqrt{\la}} 
\({\sqrt{2}\over\La \sqrt{{\rm Vol}_W} R^{3/2}}\)^{-4}
{\cal R}(u).
\end{equation}
We have divided out the leading 
dependence on $s$ and the dependence on the hadron wave 
function normalizations. We have also divided out a
factor of $e^{4u/\sqrt{\lambda}}$, which might seem
odd since this has explicit $u$
dependence. However, this factor is 1 at small $u$ (since
$\lambda \gg 1$) and irrelevant at large $u$ (since the glueball wave 
function falls exponentially at large $u$). The definition (\ref{dimlessR}) 
is convenient in that it makes the function $\widehat{\cal R}$
positive-definite and independent of $\lambda$, except through
the dependence on the rapidity.

Figure \ref{fig:R_t0} shows $\widehat{\cal R}$ at
three different values of $y$; the left plot is for $\tau=1$ and the
right plot for $\tau=10$. The peak, whose position in $u$ 
scales as $\sqrt{\tau}$, and the ensuing
Gaussian fall-off (\ref{gaussianfalloff}) beyond the peak, are
clearly visible.
As $\tau$ increases, the height of the peak decreases,
consistent with the $(\tau_1 \tau_2)^{-3/2}$ scaling of the amplitude.
However, the relative height of the peak compared to the value of
$\widehat{\cal R}(0)$ increases.  Finally,
as we change $y$ at fixed $\tau$, the shape of $\widehat{\cal
R}(u)$ changes little for $|y| \ll \tau\sqrt{\lambda}/2$.

\begin{FIGURE}[h]{
\centerline{\psfig{figure=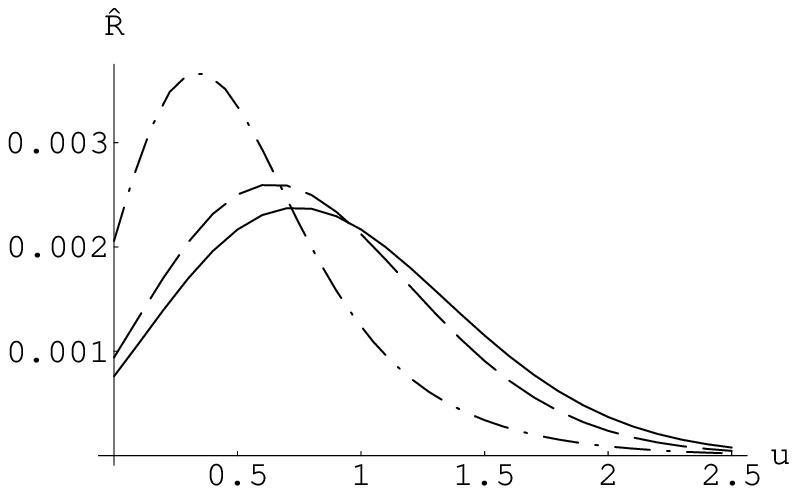,width=2.5in}
~~~~
\psfig{figure=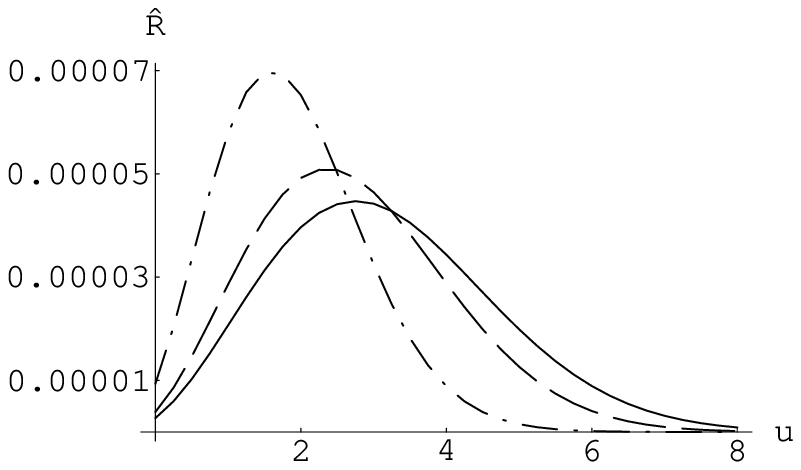,width=2.5in}
}
\vspace*{-20pt}
\caption{The dimensionless double Pomeron source function
$\widehat{\cal R}$, defined in eq.~(\ref{dimlessR}), plotted versus
the $AdS$ radial coordinate $u$. Each plot is at a fixed value of $\t$
and exhibits $\widehat{\cal R}$ at different rapidities.  The left
plot has $\t = 1$ and shows curves for $y = 0$ (solid),
$0.5\sqrt{\la}$ (dashed), and $0.9\sqrt{\la}$ (dot-dashed). The right
plot has $\t = 10$ and shows curves for $y = 0$ (solid), $5\sqrt{\la}$
(dashed), and $8\sqrt{\la}$ (dot-dashed).
For the external hadrons, we have chosen $\D_0 = 3$, $m_0/\La =
2.405...$. Since $\lambda\gg 1$, there is a large range in rapidity over
which $\widehat{\cal R}$ is nearly rapidity-independent.}
\label{fig:R_t0}
}
\end{FIGURE}

\subsection{Finite $t_1, t_2$ and large $\tau_i$: absence of a Regge
peak and power-law behavior}
\label{sec:finitetlargetau}
The scattering amplitude for our diffractive process is peaked at
$t_i=0$ (within the physical regime).  In classic Regge physics, one
might expect an exponential fall-off with negative $t_i$.  We will see
in the next section that this is true, but only at small $\tau_i$. 
In this section we will show that at
large $\t_i$, and for negative $t_i$, the amplitude decreases like a
power of $t_i$, as in \cite{HS}.  

First, we show Regge behavior is absent at small negative
$t_i$.
Starting with (\ref{Smatrix}), we calculate the amplitude 
as in section 7.1. In particular, we must
compute integrals similar to those 
in (\ref{P}) and (\ref{Q}), where there is now additional 
dependence on the $t_i$.  
We can only compute
$P$ and $Q$ analytically when $\t, \t_1, \t_2$ are large 
enough that the Gaussian factors rapidly cut off
the integrals over $\nu$. 
The eigenfunctions of (\ref{diffusiveH}) for all values of $t_i$ in
the hard-wall model are given in appendix \ref{sec:eigfuncs}.  We expand them
to lowest order in $\nu$, to find
\begin{eqnarray}
\label{negtpsisq}
\psi_\nu^*(t_i,u)\psi_\nu(t_i,u') &=& {2\over\pi}\nu^2
[K_0(\rho_i e^{-u}) + H(\rho_i)I_0(\rho_i e^{-u})] \times \\
&& \qquad \qquad \qquad \qquad
[K_0(\rho_i e^{-u'}) + H(\rho_i)I_0(\rho_i e^{-u'})] + O(\nu^4), \nonumber
\end{eqnarray}
where $\rho_i \equiv \sqrt{-t_i}/\La$ and
\begin{equation}
H(\rho_i) = {-2K_0(\rho_i) + \rho_i K_1(\rho_i)\over
2I_0(\rho_i) + \rho_i I_1(\rho_i)}.
\end{equation}
$H(\rho_i)$ vanishes exponentially as $\rho_i$ approaches infinity; it diverges
logarithmically when $\rho_i$ goes to 0, though $\psi_{\nu}(0,u)$ is finite.

We approximate (\ref{D0b}) as
\begin{equation}
\label{D0largeti}
D_0(u,u',u'') 
\approx  
2\sqrt{\la}\int d\nu\, \psi_\nu(t_1,u)\psi_\nu^*(t_1,u')
\int d\w\, \psi_\w(t_2,u)\psi_\w^*(t_2,u'') 
e^{-\t_1\nu^2 -\t_2\w^2}
\end{equation}
where we have taken $\tau_\perp$ to be negligible and kept the
$\nu$ and $\omega$ dependence only in the exponentials where
they are multiplied by the large numbers $\tau_i$.
The kernel is now easily computed,
\begin{eqnarray}
D_0(u,u',u'') & \approx & {\sqrt{\la}\over 2\pi(\t_1\t_2)^{3/2}}
[K_0(\rho_1 e^{-u}) + H(\rho_1)I_0(\rho_1 e^{-u})]
[K_0(\rho_1 e^{-u'}) + H(\rho_1)I_0(\rho_1 e^{-u'})] \nonumber \\
& & \times
[K_0(\rho_2 e^{-u}) + H(\rho_2)I_0(\rho_2 e^{-u})]
[K_0(\rho_2 e^{-u''}) + H(\rho_2)I_0(\rho_2 e^{-u''})]. \label{negtkernel}
\end{eqnarray}
Notice that the dependence on $u$, $u'$, $u''$ completely factorizes.
Also, the dependence on $\tau$ and $\tau_i$ completely factorizes from
the dependence on $t_i$.  {\it This second factorization implies that there
is no Regge behavior $s_i^{\alpha(t_i)} \sim e^{\alpha(t_i) \tau_i}$
in the large $\tau$ and $\tau_i$ limit.}  Looking back, we see this
factorization of the $t_i$ and $\tau_i$ dependence stems from the
factorization of the $\nu$ and $\rho_i$ dependence in
eq.~({\ref{negtpsisq}}).  As long as $\psi^*_\nu(t_i,u)\psi_\nu(t_i,u')$ has
a power series expansion in $\nu$ near $\nu=0$, this factorization,
and the corresponding loss of Regge behavior, will always occur at
large $\tau_i$, for any large-$\lambda$ theory. 

Before continuing we should note a subtlety.  The $D_0$ kernel
appearing in the above expression is unbounded as any of $u$, $u'$,
$u''$ increase, which might seem problematic.  However, in deriving
this expression we used an approximation which breaks down at large
$u$, $u'$ or $u''$.  One can see explicitly that when $t_i=0$,
the corresponding expression (\ref{D0approx}) is strongly damped
at large $u$, $u'$, $u''$ by Gaussian factors. 
Here, our use of an expansion in $\nu$, assuming the
$\tau_i$ are large, is essentially (after integrating over
$\nu$) an expansion in $u^2/\tau$.  This is not valid where
the Gaussian factors are important, so $D_0$ in this regime
has no large-$u$ cutoff.  However, the hadron and glueball
wave functions, which are integrated against $D_0$ to obtain the source
function and the S-matrix, have their own exponential tails in
$u$.  These tails cut off the integrals at $u\sim 1$, long before our
approximation breaks down.  We therefore proceed without concern.

Next, we show that the scattering amplitude falls off as a power law in
$t_i/\Lambda^2$, another universal phenomenon.  The presence of falling
powers is easy to understand.  At large negative values of $t_i$, the
effective potential in the Hamiltonian (\ref{diffusiveH}) develops an
exponential barrier that forces the corresponding analogue particle away
from the wall.  In the calculation of the scattering amplitude, the
eigenfunctions of this Hamiltonian are integrated against the
wave functions of the glueball and external hadrons, which have falling
power-law tails in $z$ at small $z$ (exponentially falling tails in
$u$ at large $u$).  For large values of $t_i$, the eigenfunctions lack
support near the wall, and the computation is dominated by the
power-law tails of the wave functions.  Once this is true, the entire
computation scales with $t_i$, giving the amplitude a power-law
dependence on $t_i$.

To demonstrate this explicitly, we insert (\ref{negtkernel}) into
(\ref{Pomeron}). We are left with two factorized integrals over $u'$ 
and $u''$, which we will denote $I'$ and $I''$.  In the limit of large 
$\rho_i$, $H(\rho_i)$ vanishes, so the integral $I'$ 
becomes
\begin{equation}
I' \approx \int_0^\infty du'\, \phi_0(u')^2 K_0(\rho_1 e^{-u'})
= \int_0^{z_{0}} \ \frac{dz'}{z'}\, \phi_0(z')^2 K_0(\sqrt{|t_1|}z').
\end{equation}
The integral $I''$ is identical with $t_2$ replacing $t_1$.  
Since the function $K_0(x)$ is exponentially damped for large $x$,
the integrand only has support for $z' \ll 1/\sqrt{|t_1|}$.  
In this region, the external hadron wave function 
$\phi_0(z')$ has a power-law tail: 
\begin{equation}
%
\phi_0(z') \approx \frac{\sqrt{2}\Lambda}{ \sqrt{\rm{Vol}_W} R^{3/2}} 
\frac{(m_0/2)^{\Delta_0-2}}{\Gamma(\Delta_0 - 1) 
J_{\Delta_0-2} (m_0/\Lambda)}z'^{\Delta_0} ~~~~ (z'\ll m_0^{-1} \sim z_0).
\end{equation}
(Recall that, for low-lying external hadron states, 
$m_0 \sim \Lambda = 1/z_{0}$.)  
With this approximation, $I'$ can be evaluated by 
extending the range of integration over
$z'$ from zero to infinity.
Inserting the integrals $I'$ and $I''$ into the definition 
of $\mathcal{R}(u)$, eq.~(\ref{Pomeron}), gives
\begin{equation}
 \mathcal{R}(u)  \approx  \frac { 2 \chi (\Delta_0, m_0) } 
{ \pi {\rm Vol}_W^2 R^6 \sqrt{\lambda} } 
\( e^{-i\pi/2} \frac{ s} { 4 \Lambda^2 } \) ^{2 - 2/\sqrt{\lambda}} 
\( \frac { m_0^4 }{ t_1 t_2}\)^{ \Delta_0} 
\frac {e^{4u/\sqrt{\lambda} } } {(\t_1 \t_2)^{3/2}}
 K_0 (\rho_1 e^{-u})
 K_0 (\rho_2 e^{-u})
\label{largetR}
\end{equation}
where 
\begin{equation}
\chi(\Delta_0, m_0) = \left( \frac { 2 (\Delta_0 - 1) } 
{  J_{\Delta_0 - 2} (m_0/\Lambda) }  \frac {\Lambda^2} {m_0^2} \right)^4 \ .
%
\end{equation}
This expression is valid in the region of our approximation, $u\ll
\sqrt{\tau}$, as noted above.  
Note the almost-quadratic growth of the amplitude with $s$, the
power-law dependence on the $t_i$, and the $(\tau_1\tau_2)^{-3/2}$
factor, which provides subleading $s$ dependence and limited rapidity
dependence.  The two Bessel functions determine the shape of the
function in $u$ and contribute a subleading dependence on the $t_i$.
Since $\rho_i=\sqrt{|t_i|}/\Lambda\gg 1$, these functions
assure an exponential cutoff of the source function near the wall at $u=0$.
The growth of the source function as
$u\to\infty$ is cut off by the breakdown in our approximation.  But before
this happens, the growth is more than compensated by a falling
glueball wave function $(\phi_5(z)\sim z^4)$ in our current
computation.

We can now compute glueball production at large $t_i$, using the 
above Pomeron source function. 
The  amplitude involves the integral of the glueball 
wave function against the Pomeron source function.  The 
factors of $K_0(\sqrt{|t_i|} z)$ are damped exponentially near the wall,
so the glueball wave function can be approximated by its power-law tail.
Also, the factor $e^{4u/\sqrt{\lambda}}$ can be replaced by 1.  

The scattering amplitude (\ref{amplitude}) obtained from (\ref{SRrule}) is
\begin{equation}
\label{T4def}
{\cal T}_4 = -i\frac{{\cal N}\a'^5 g_s^3}{\sqrt{2}\pi{\rm Vol}_W^{3/2}
R^{5/2}\sqrt{\la}(\t_1\t_2)^{3/2}\La}
\(e^{-i\pi/2}\frac{s}{4\La^2}\)^{2-2/\sqrt{\la}}\Omega(t_1,t_2)
\end{equation}
where
\begin{equation}
\Omega(t_1,t_2) \approx \frac{ 2\chi(\Delta_0, m_0)} {J_2(m/\Lambda)} \(
\frac{m}{\Lambda}\)^2 \( \frac { m_0^4 }{ t_1 t_2 }\)^{ \Delta_0}
\frac{\Lambda^4}{(t_1-t_2)^2} \[ -1 +
\frac{1}{2}\frac{t_1+t_2}{t_1-t_2} \ln \(t_1\over t_2\) \] .
\end{equation}
The amplitude is well-behaved as $t_2 \rightarrow t_1$, 
since
\begin{equation}\label{T4approx2}
\lim_{t_2 \to t_1} \Omega = 
\frac{\chi(\Delta_0, m_0)} {6 J_2(m/\Lambda)} 
\( \frac{m}{\Lambda}\)^2 
\frac{m_0^{4\Delta_0} \Lambda^4}{t_1^{2\Delta_0+2}}.
\end{equation}
For equal momentum transfers, the scaling of our amplitude in 
terms of gauge theory variables is simply
\begin{equation}
\label{T4scaling}
{\cal T}_4 \sim 
\frac{1}{\La N^3}\frac{s^{2-2/\sqrt{\la}}}{t_1^{2\Delta_0+2}}
(\tau_1\tau_2)^{-3/2}.
\end{equation}

Although we are not quite in the same limit, we may compare
(\ref{T4scaling}) with the high energy, fixed-angle 
scattering result of ref.~\cite{HS}.  At large $s$ and $t_i$ but
fixed $s/t_i$, eq.~(14) of ref.~\cite{HS} indicates that in $2\to 3$ scattering
of four objects created by operators of dimension $\Delta_0$ and one of
dimension  $\Delta=4$,
the amplitude would scale as
\begin{equation}
\label{HScompare}
{\cal T}_4 \sim \frac{\la^{\Delta_0+1/2}}{\La N^3}
\(\frac{\La}{p}\)^{4\Delta_0}
\end{equation}
where $p$ is a characteristic momentum scale for the $2 \to 3$
scattering process.  General requirements assure the overall power of
$g_s \sim 1/N$ is the same in both cases as well as the $1/\Lambda$ sitting in
front.  For low-lying hadrons $m\sim m_0\sim \Lambda$.  Our amplitude
scales as $s^2 t_i^{- 2\Delta_0-2}$, exhibiting a power law
suppression in momentum with the same overall exponent as
(\ref{HScompare}); both scale as $(\Lambda/p)^{4 \Delta_0}$.
However, the dependence on $\sqrt{\lambda} = R^2 / \alpha'$ in the two
expressions is different.  This is because the dynamics of string
theory itself --- the exponential suppression of hard scattering ---
cuts off the amplitude in \cite{HS}, which introduces factors of
$\alpha'$.  Here the cutoff on the amplitude occurs through the $K_0$
Bessel functions above, through the large momentum transfers $t_1$ and
$t_2$.  These are independent of $\alpha'$.

\subsection{Small $t_1$, $t_2$ and small $\tau_i$: classic Regge phenomenology}

In the classic Regge regime of $s \gg |t| \sim \Lambda^2$, experiments
show hadronic amplitudes exhibit Regge behavior, $\mathcal{S} \sim
s^{\alpha(t)}$ with $\alpha(t)$ approximately linear in $t$.  We have
just seen that for large $\tau_i$ this behavior is absent.  What has
happened?  In \cite{HS} it was argued that for fixed $t$ the
classic Regge behavior of the amplitude is a transient phenomenon
present only for a certain range of $s$.  When $s$ is sufficiently
large, a transition takes place and power-law behavior is restored.
Here we will see a similar phenomenon.
In this section, we show that for small enough $t_i$ and
$\tau_i$, one finds approximate Regge behavior, as a transient
effect that gives way, for larger $\tau_i$, to the power-law
behavior seen in the previous section.  More specifically, we will work
in the regime
\begin{equation}
\label{window}
\frac{1}{\sqrt{\lambda}} \ll \tau_i \ll 1 \qquad \mbox{and}
\qquad
\frac{\tau_i |t_i|}{\Lambda^2} \sim 1 \ .
\end{equation}
The lower bound on $\tau_i$ is the double Regge limit we have taken
from the beginning.  The upper bound on $\tau_i$ keeps the diffusion
times short enough that the potential (\ref{V}) can be approximated by
its first few terms in a Taylor series expansion.  The second relation
keeps the product $\tau_i |t_i|$ large enough that we can distinguish
between exponential and linear behavior of the amplitude as a function
of $t_i$.
We will argue that the presence of this window, where Regge scaling
can be seen, is universal and is not sensitive to the details of the
hard-wall model.

The generality of the phenomenon can be seen from the following
argument.  The source function $\mathcal{R}(u)$ involves the combination of
two Pomerons, one from each of the external hadrons, 
\begin{equation}
\label{Rpomeron}
\mathcal{R}(u) \sim 
\langle u| e^{-H_1\tau_1} | \phi_0^2 \rangle \langle u| e^{-H_2\tau_2} | 
\phi_0^2 \rangle \ ,
\end{equation}
each of which
requires a diffusion-kernel calculation of the matrix element
\begin{equation}
\label{matrix}
\langle u| e^{-H_i\tau_i} | \phi_0^2 \rangle = 
\int du' \langle u|
e^{-H_i\tau_i} | u' \rangle \langle u'| \phi_0^2 \rangle
\end{equation}
where $H_i$ is the Hamiltonian (\ref{diffusiveH}) and 
$|\phi_0^2\rangle$ is the state whose wave function is $\phi_0^2(u)$
in the position basis.  
That $\mathcal{R}(u)$ is given by (\ref{Rpomeron}) can
be seen from examination of (\ref{Pomeron}) and (\ref{D0b}).

For small $\tau_i$ and $t_i$ we may evaluate
the matrix element (\ref{matrix}) semiclassically, 
writing it as a path integral over
all paths between $u$ and $u'$, and expanding around the classical
path of minimal Euclidean action.  Defining the average position
$\bar{u} = (u+u')/2$ we may rewrite the potential as
\begin{equation}
V_i(u') \sim 4 - z_0^2 t_i e^{-2\bar{u}}[1 -2(u' - \bar{u}) + 2(u' -
\bar{u})^2 + ... ] \ .  
\end{equation}
Working to linear order in the potential, the matrix element is then
\begin{equation}
\langle u| e^{-H_i\tau_i} | \phi_0^2 \rangle = 
\int du' \exp\left[z_0^2 t_i \tau_i e^{-2\bar u}\right]
e^{-4\tau_i} 
\phi_0^2(u') \int \mathcal{D}
y ~ e^{-\int_0^{\tau_i} ds [\frac{1}{4} \dot{y}^2 +
\frac{1}{2}\beta(y-\bar{u})]} .
\end{equation}
Here 
$\beta = 4 z_0^2 t_i e^{-2\bar{u}}$, and the boundary conditions on
the path integral are that $y(0) = u'$ and $y(\tau_i) = u$.  The
classical solution is $y_{cl}(s) = \frac{1}{2}\beta s^2 +v_0 s +u'$
where $v_0 = -\frac{1}{2} \beta \tau_i + \frac{u-u'}{\tau_i}$.
Letting $y = y_{cl} + x$, the path integral becomes\footnote{
 Keeping the quadratic term in the potential $V_i(u')$ will produce
 contributions to the transition amplitude that are subleading in 
 $\tau_i$.  One might worry that it is inconsistent to consider 
 the quadratic $\dot x^2$ fluctuations without the $x^2$ fluctuations.
 Keeping the $x^2$ fluctuations, the quantum determinant
 scales as $|\beta_i|^{1/4} \mbox{csch}(2 \sqrt{2|\beta_i|} \tau_i)^{1/2}$
 which is indeed independent of $\beta_i$ for small $\tau_i$
 (see for example \cite{Cohen}).
}

\begin{equation}
\langle u| e^{-H_i\tau_i} | \phi_0^2 \rangle =  \int du'  
\exp\left[z_0^2 t_i\tau_i e^{-2\bar u}\right] e^{-4\tau_i
+\frac{\beta^2 \tau_i^3}{48}  - \frac{(u-u')^2}{4\tau_i}} 
\phi_0^2(u') 
\int \mathcal{D} x ~ e^{-\int_0^{\tau_i} \frac{1}{4}\dot{x}^2 ds},
\end{equation}
where the boundary conditions on $x$ are $x(0) = x(\tau_i) = 0$.  The
remaining path integral is proportional to $1/\sqrt{\tau_i}$ but
is independent of $t_i$ and plays no further role.
The Gaussian factor in the $u'$ integral, along with the constraint
on $\tau_i$ (\ref{window}),
constrains $u'$ to be near $u$.  
Expanding the rest of 
the integrand around $u' = u$, performing the Gaussian integral, 
and dropping terms
suppressed explicitly by powers of $\tau_i$, we obtain, 
after substituting the explicit form for $\phi_0(u)$,
\begin{equation} 
\label{LeadingPomeron}
\langle u| e^{-H_i\tau_i} | \phi_0^2 \rangle \sim 
e^{-4u -b_ie^{-2u} 
}
J_{\Delta_0-2}([m_0/\Lambda] e^{-u})^2 
\mathrm{erfc}\left(
\frac{  - u}{2 \sqrt{\tau_i}} 
\right) + O(\sqrt{\tau_i}), 
\end{equation}
where $b_i =  - z_0^2 t_i \tau_i > 0$.  
The Pomeron source function $\mathcal{R}(u)$ is then proportional to
(\ref{LeadingPomeron}) for $i=1$ times a similar expression for $i=2$.

Finally, to obtain the glueball production amplitude, there is still
the integral over $u$ to perform, (\ref{SRrule}), of the Pomeron
source function against a glueball wave function.  That this integral
leads to approximate Regge behavior, with a Regge slope that slowly
varies with $t_i$, can be seen as follows.
The complementary error functions can be
replaced by $2$ everywhere in the range of integration except for
$u<\sqrt{\tau_i}$, a region which gives a subleading effect in
$\tau_i$.
The integral (\ref{SRrule}) is dominated by the exponential
$\exp(-b_i e^{-2u})$ and the power law tails that come from expanding
the Bessel functions at large $u$.  Thus a reasonably good approximation to
the $t_i$ dependence of the amplitude is
\begin{equation}
\int_0^\infty du\, \phi_5(u) {\mathcal R}(u)
\sim \int_0^\infty du\, e^{- b e^{-2u} - 2M u}
= \frac{1}{2} b^{-M} \left(
\Gamma(M) - \Gamma(M, b) \right) \ ,
\end{equation}
where $\Gamma(a,z) \equiv \int_z^\infty t^{a-1} e^{-t} dt$ is the
incomplete gamma function, $b\equiv b_1 + b_2$, and $M\equiv 2
\Delta_0 + 2$.  Expanding this result at large $M$ leads to
approximate Regge behavior, $e^{-b_1-b_2}$, where we recall that $b_i
\propto - z_0^2 t_i \tau_i$.

In sum, we find approximately exponential behavior in the window
(\ref{window}), and therefore Regge behavior of the amplitude, with a
Regge slope of order $1/\sqrt{\lambda}\Lambda^2$.  Once $\tau_i$ or
$t_i$ becomes large, however, this argument breaks down, Regge
behavior is lost, and the physics enters the regimes discussed in
sections \ref{sec:zerot} and \ref{sec:finitetlargetau}.

\section{Discussion}
\label{sec:discussion}

Using the hard-wall model for gauge/string duality, we studied
glueball production from hadron scattering in the double diffractive
limit.  String theory provides a new window into this 
regime, which is not accessible using perturbative QCD.
A principal objective of this paper was to set up a formalism
that can be applied to a number of $2 \to 3$ processes that are more
experimentally relevant than glueball production.
These could include heavy quarkonium production and Higgs
production.

Our study is therefore largely of technical interest, though we also
uncovered some notable physical phenomena.  We wrote our amplitudes in
terms of a source function ${\mathcal R}(u)$, which was independent of
the glueball state and characterized the two Pomerons which fuse to
make the glueball.  The amplitude is given by integrating this
function against the glueball wave function.  Applying this formalism,
we saw that Regge phenomena are only present at relatively small
$s,s_1,s_2$, and are lost as these quantities become large.  We found
that at large $s, s_i$ and small $t_i$, the scattering amplitude
becomes rapidity-independent, a fact which arises from
the rapidity-independence of the corresponding tree-level dual
gravitational amplitude.  We also found the expected power-law
fall-off of the amplitudes at large momentum transfers $t_i$, where
scaling behavior is expected, as in \cite{HS}.

A natural next step would be to apply this formalism to heavy
quarkonium production.  Quarkonium in $AdS$/CFT can be
modeled by adding a D-brane to the $AdS$ cavity that fills $AdS$ from the
boundary to some minimal radius $r_0=R^2/z_0$ \cite{KarchKatz}.  The mass of
the heavy quark scales with $r_0$, and the quarkonium supergravity
wave function must live on the D-brane.  In  principle
one ought  to consider the double Regge limit of the 
five-point disk amplitude for
string theory in flat space, with four closed string insertions, and one
open string insertion representing the heavy meson.  However, most of the
important physical effects may be captured simply by
using the fact that the quarkonium wave function has support only in the 
region $r > r_0$ $(z<z_0)$ ignoring any
further details. 
Even with such a simple model, it might be possible
to make a prediction for the relative rates of bottomonium and charmonium
production in the double diffractive limit.

Double diffractive Higgs production might be a clean way to observe
the Higgs boson (or other scalars) at the LHC.
There are competing field theory models, some inspired by flat-space
string theory, which predict potentially observable cross-sections for
exclusive Higgs production at the LHC.  See ref.~\cite{Forshawreview}
for a brief review.  It is possible that gauge/string duality might
clarify some of the approximations made in these models.  More
concretely, we would proceed as follows.  The Higgs, which couples to
$F_{\mu\nu}F^{\mu\nu}$ in the standard model through a dimension-five
operator, should be treated similarly to the scalar glueball in our
discussion above, except for one crucial difference: we should replace
the normalizable dilaton mode in supergravity, representing the
glueball, with its non-normalizable counterpart, representing the
Higgs boson. This non-normalizable mode, at timelike momentum $p^\mu$
with $-p^2=m_H^2$, is highly oscillatory.  The technical challenge
that lies ahead is then to compute the fusion of two Pomerons into
such a state.  We expect that this challenge can be met, and that
gauge/string duality will soon contribute to the debate over
diffractive Higgs production.

\section*{Acknowledgments}

We would like to thank Andreas Karch, Richard Brower and Chung-I Tan
for discussions. The work of M.J.S. was supported in part by the 
U.S. Department of Energy under Grant No. DE-FG02-96ER40956.
The work of C.P.H. was supported in part by the National Science 
Foundation under Grant No. PHY-0243680.
The work of S.P. and E.G.T. was supported in part by a Royalty Research Fund
award from the University of Washington.

\appendix

\section{Bosonic tachyon/dilaton amplitude}
\label{sec:fivepointcalc}
In this appendix, we calculate the double diffractive limit of
the four tachyon, one dilaton, tree level,
closed string amplitude in bosonic string theory in 
twenty-six dimensional Minkowski spacetime.  
We let the dilaton correspond to the fifth particle in Figure \ref{fig:2to3}.
Following closely an approach described in \cite{Lipatov:1987nn},
we are able to express the result both as a power
series in $\alpha' m_\perp^2$ and an asymptotic series in 
$1/\alpha' m_\perp^2$.  The power series result is important
for our discussion of glueball production in the paper; 
for glueball production,
$\alpha' m_\perp^2$ is effectively very small.

We use the conventions in Polchinski's textbooks
\cite{Polchinski}. In section 
\ref{sec:Smatrix}, string theory momenta have a $\widetilde{~}\,$ 
to distinguish them from four-dimensional gauge theory 
momenta. We will not need to make that distinction here as we do
not work with the four-dimensional gauge theory momenta at all.
Unlike the conventions in the main body of the paper, we take all of the 
momenta $k_i$ to be ingoing.

Take four tachyon vertex operators $T_i = g_c \, {:}e^{ik_i\cdot X}{:}$, 
$i = 1,\ldots,4$ with $k_i^2 = 4/\a'$, and a dilaton vertex 
operator $D = g'_c f_{\mu\nu}\, {:}\p X^\mu \pbar X^\nu e^{ik_5\cdot X}{:}$ 
with $k_5^2 = 0$. The dilaton is a massless particle that travels 
on lightlike geodesics so the symmetric tensor $f_{\mu\nu}$ must 
be transverse to its momentum, $k_5^\mu f_{\mu\nu} = 0$. Given 
another lightlike vector $\kbar$ such that $k_5\cdot \kbar \neq 0$, 
we can satisfy the constraint by taking 
$f_{\mu\nu} = \eta_{\mu\nu} - (k_{5\mu}\kbar_\nu + 
\kbar_\mu k_{5\nu})/k_5\cdot \kbar$. 
The correlation function of 
these vertex operators on the Riemann sphere is
\begin{eqnarray}
\label{correlator}
\EV{\prod_{i=1}^4 T_i(z_i,\zbar_i) D(z_5,\zbar_5)} & = & 
iC_{S^2}^X g_c^4 g_c' 
(2\pi)^{26}\d^{26} \(\sum k_i \) \prod_{1\leq i < j \leq 5} 
|z_{ij}|^{\a'k_i\cdot k_j} \cr
& & \times 
f_{\mu\nu} \(-i{\a'\over2}\sum_{i=1}^4{k_i^\mu\over z_{5i}}\)
\(-i{\a'\over2}\sum_{j=1}^4{k_j^\nu\over \zbar_{5j}}\).
\end{eqnarray}
We have used the shorthand notation $z_{ij} \equiv z_i - z_j$. 

The tree-level S-matrix for the scattering of four tachyons and a 
dilaton is obtained by integrating the correlator (\ref{correlator}) 
over all possible worldsheet coordinates for the operator insertions, 
weighted by a topological factor,
\begin{equation}
{\cal S} = e^{-2\la} 
\prod_{i=1}^5 
\int_{\mathbb{C}\cup\{\infty\}} 
d^2z_i\, \D_{\rm ghost} \EV{\prod_{i=1}^4 T_i(z_i,\zbar_i)
D(z_5,\zbar_5)}.
\end{equation}
The path integral over ghost fields contributes a Jacobian 
$\D_{\rm ghost} = C_{S^2}^g \d^2(z_a-z_a^0)\d^2(z_b-z_b^0)
\d^2(z_c-z_c^0)|z_{ab}z_{bc}z_{ca}|^2$ that fixes the residual 
$PSL(2,\mathbb{C})$ symmetry left over from conformal gauge-fixing. 
This gives us freedom to fix three vertex operators to arbitrary 
positions. A convenient choice is $z_1 = 0, z_4 = 1, z_5 = \infty$, 
because that makes $|z_{14}z_{45}z_{51}|^2 = |z_5|^4$ and 
$\prod_{i<5}|z_{i5}|^{\a'k_i\cdot k_5} = 1$. Since (\ref{correlator}) 
comes with a factor of $|z_5|^{-4}$, the S-matrix is finite.\footnote{%
Note, if $z=x+iy$, we define the measure factor $d^2z \equiv 2 dx \, dy$.
}
\begin{equation}
{\cal T}_{26} = C\int d^2z_2 d^2z_3 \prod_{1\leq i < j \leq 4} 
|z_{ij}|^{\a'k_i\cdot k_j}
f_{\mu\nu} (k_2 z_2 + k_3 z_3 + k_4)^\mu 
(k_2 \zbar_2 + k_3 \zbar_3 + k_4)^\nu.
\end{equation}
We have from ref.~\cite{Polchinski} that 
$e^{-2\la}C_{S^2}^XC_{S^2}^g = 8\pi/\a'g_c^2$ and 
$g_c' = 2g_c/\a'$,
which leads to the result that $C = -4 \pi g_c^3$.
In these conventions, the coupling
$2\pi g_c = \ka$, where $\ka$ is the gravitational coupling constant
that appears in front of the Einstein-Hilbert action as $(2\ka^2)^{-1}$.
Thus, we find that $g_c \sim \alpha'^{(d-2)/4} g_s$
where $g_s$ is the dimensionless string coupling
(which is related to the vev of the dilaton)
and $d$ is the number of dimensions in which we work.

Make the conformal transformation $u = z_2/z_3$ and 
$v = (z_3-1)/(z_2-1)$ to get
\begin{eqnarray}
{\cal T}_{26} & = & C\int d^2u \, d^2v 
|u|^{-\a't_1/2-4} |v|^{-\a't_2/2-4}
|1-u|^{-\a's_1/2-4} |1-v|^{-\a's_2/2-4} |1-uv|^{\a'(-s+s_1+s_2)/2} \nonumber\\
& & \times f_{\mu\nu} \left[k_2u(1-v) + k_3(1-v) + k_4(1-uv)\right]^\mu
[{\rm c.c.}]^\nu.
\end{eqnarray}
Now take the double Regge limit given in (\ref{ddlimit}). 
The dominant integration regions are near the origins of the $u$ and
$v$ planes. In order to demonstrate this we should first discuss the
issue of convergence. For fixed $v$ and for real Mandelstam
variables with physical signs, there are clearly three
special points in the finite $u$-plane where the integrand becomes singular,
$u = 0, 1, v^{-1}$. Of these, the origin is benign as long as we take 
$\a't_1 < -4$ so that the singularity is integrable. The other singularities
at $u=1,v^{-1}$ are not integrable for positive $s_1$ and $s$. One way
to avoid this difficulty is to choose pure imaginary values for the 
$s$-type Mandelstam variables and analytically continue the result
to the physical domain. In this scheme $|1-u|^{-\a's_1/2}$, 
$|1-v|^{-\a's_2}$ and $|1-uv|^{\a'(-s+s_1+s_2)/2}$ are just phases.
Now if $\a't_1 > -8$, then the integrand vanishes sufficiently fast at 
infinity for the $u$-plane integral to converge. Thus, the entire integral may
be defined by continuation from $-4 > \a't_i > -8$ and pure imaginary
$s, s_1, s_2$. We observe that this range of momentum transfers
implies that $|u|^{-\a't_1/2-4}$ and $|v|^{-\a't_2/2-4}$ are always 
singular near the origins of the $u$ and $v$ planes.

For fixed $v$, the integral in $u$ is dominated by a saddle point at 
$u \sim O(1/s_1)$, which approaches the origin in the Regge limit. 
Therefore, $|1-u|^{-\a's_1/2-4} \goesto e^{\a's_1(u+\ubar)/4}$. 
Similarly, for fixed $u$, the integral in $v$ is dominated by a saddle point 
at $v \sim O(1/s_2)$. It follows that $|1-v|^{-\a's_2/2-4} \goesto 
e^{\a's_2(v+\vbar)/4}$ and $|1-uv|^{\a'(-s+s_1+s_2)/2} \goesto 
e^{\a's(uv+\ubar\vbar)/4}$. Therefore,
\begin{eqnarray}
\label{Tregge}
{\cal T}_{26} & \approx & C\int d^2u \, d^2v
|u|^{-\a't_1/2-4}|v|^{-\a't_2/2-4} 
\exp\left[\a's_1\Re(u)/2 + \a's_2\Re(v)/2 + \a's \, \Re(uv)/2\right] 
\nonumber\\
& & \times f_{\mu\nu}\left[k_2 u (1-v)+k_3(1-v)+k_4(1-uv)\right]^\mu 
[{\rm c.c.}]^\nu  \ .
\end{eqnarray}

We still need to understand the consequences that the Regge limit has
on the second line of (\ref{Tregge}).
It suffices to consider only the $\eta_{\mu\nu}$ part of the 
tensor $f_{\mu\nu}$ since the portion containing $\kbar$ 
corresponds to a longitudinal polarization that must decouple in
physical processes \cite{GSW}. Moreover, $\kbar$ is arbitrary 
and the final amplitude cannot depend on it.  So
\begin{eqnarray}
\label{contracs}
\lefteqn{ 
\eta_{\mu\nu}\left[k_2 u (1-v)+k_3(1-v)+k_4(1-uv)\right]^\mu[{\rm c.c.}]^\nu }
\nonumber\\
& = & 
(k_3+k_4)^2 + k_2\cdot(k_3+ k_4)(u+\ubar) - k_3\cdot(k_3+k_4)(v+\vbar) 
\nonumber\\
& & + ~ ({\rm terms~with~two~or~more~factors~of~} u, \ubar, v, \vbar).
\end{eqnarray}
Scaling $u \goesto u/s_1$ and $v \goesto v/s_2$ in 
(\ref{Tregge}), it is clear that only the first three terms on the
right hand side of (\ref{contracs}) have the
possibility of not being suppressed by a large energy.\footnote{A 
careful treatment of this scaling is given later in this appendix.}
Since $k_3\cdot(k_3 + k_4) = -t_2/2$, the third term is actually
suppressed. Using $k_2\cdot(k_3+k_4) = (t_2-s_1-4/\a')/2$, 
we may replace (\ref{contracs}) by $-t_2 - s_1(u+\ubar)/2$. 

For convenience we define a prototypical integral
\begin{equation}
\label{proto}
I(a,\abar,b,\bbar) = \int d^2u \, d^2v\, u^a \ubar^{\abar} v^b \vbar^{\bbar} 
\exp\left[\a's_1\Re(u)/2 + \a's_2\Re(v)/2 + \a's \, \Re(uv)/2\right].
\end{equation}
With $a = -\a't_1/4-2$ and $b = -\a't_2/4-2$, the amplitude reads
\begin{equation}
\label{Tproto}
{\cal T}_{26}/C \approx -t_2 I(a,a,b,b) 
- {s_1\over 2}[I(a+1,a,b,b) + I(a,a+1,b,b)].
\end{equation}
${\cal T}_{26}$ is a function of five variables $s, s_1, s_2, t_1, t_2$. 
It is still not well-defined for physical scattering (real $s > s_1 + s_2 > 0$ 
and real $t_1, t_2 < 0$) since $I$ diverges. For a proper definition, 
we extend $s, s_1, s_2$ to the complex numbers and follow a 
technique due to Lipatov \cite{Lipatov:1987nn}: decompose 
$I$ over different regions 
of the $u$ and $v$ planes, and for each region choose the phases of 
$s, s_1, s_2$ such that they lie on the real axes with signs that 
ensure the convergence of the integral. ${\cal T}_{26}$ is then
defined by analytic continuation of $s, s_1, s_2$ to the positive real axes.

We now carry out Lipatov's procedure for $I$ as given in 
(\ref{proto}). It is convenient to let $\a' = 4$. At any 
fixed value of $(v,\vbar)$ the integrand has a saddle point 
at $(u,\ubar) = (-a/(s_1+sv),-\abar/(s_1+s\vbar))$. Similarly, for any 
fixed value of $(u,\ubar)$ the integrand has a saddle point at 
$(v,\vbar) = (-b/(s_2+su),-\bbar/(s_2+s\ubar))$. These saddle points 
lie close to the origin of the complex plane. 

We begin by evaluating (\ref{proto}) directly which yields
an asymptotic series in $(\alpha' m_\perp^2)^{-1}$.  
Such an expansion is useful in the regime $\alpha' m_\perp^2 \gg 1$,
but the glueball production process investigated in the paper involves
the opposite regime $\alpha' m_\perp^2 \ll 1$.
Thus, we follow by changing the variables of integration 
in (\ref{proto}) to yield
an integral over $uv$ and a power series in $\alpha' m_\perp^2$.

\subsection{Large $\alpha' m_\perp^2$}

We divide up (\ref{proto}) into four pieces, depending on whether $u$ and $v$
have positive or negative real parts,
\begin{equation}
I = \sum I^{\sigma_u, \sigma_v} \ ,
\end{equation}
where we have introduced the notation $\sigma_x \equiv \sgn\, \Re(x)$.
In each domain, we choose $s_1$ and $s_2$ to have opposite signs
from $\Re(u)$ and $\Re(v)$, respectively.  In this way, the exponential
in (\ref{proto}) damps, ensuring that the integral converges.

For $I^{\sigma_u,\sigma_v}$, we choose $s_1 = |s_1|e^{i\pi(\s_u+1)/2}$
and $s_2 = |s_2|e^{i\pi(\s_v+1)/2}$ and change variables to 
$w = e^{-i\pi} s_1 u$, $\overline{w} = e^{-i\s_u\pi}
s_1 \overline{u}$, $z = e^{-i\pi} s_2 v$, and $\overline{z} =
e^{-i\s_v\pi} s_2 \overline{v}$. These peculiar
transformations guarantee that $\wbar$ is the complex conjugate of 
$w$ and that $\zbar$ is the complex conjugate of $z$. Then we find
\begin{equation}
 I^{\sigma_u, \sigma_v} = (e^{-i\pi} s_1)^{-a-1} 
(e^{-i \sigma_u \pi} s_1)^{-\overline a-1}
 (e^{-i\pi} s_2)^{-b-1} (e^{-i \sigma_v \pi} s_2)^{- \overline b-1} J_L
 \end{equation}
where
 \begin{equation}
 J_L \equiv \int\limits_{\substack{\Re(w)>0 \\ \Re(z) > 0}}
 d^2 w \, d^2 z \, w^a \overline w^{\overline a} z^b \overline z^{\overline b}
 \exp\left[-(w+\overline w) -(z + \overline z) +
 \frac{s}{s_1 s_2} (wz + \overline {wz}) \right] \ .
\label{JL}
\end{equation}

Now $J_L$ depends only on the combination $s/s_1 s_2$.  Restoring 
$\alpha'$, this is the dimensionless ratio $4 / \alpha' m_\perp^2$.  In
the limit where $\alpha' m_\perp^2 \gg 1$, we may expand the exponentials
in (\ref{JL}) as a double power series,
\begin{equation}
J_L = \sum_{n,m\geq 0} \frac{(\alpha' m_\perp^2/4)^{-n-m}}{n!m!}
K(a+n, \overline{a} + m) K(b+n, \overline{b} + m)
\end{equation}
where\footnote{After changing to polar coordinates and rescaling, 
the integral $K$ factors into the product of two one-dimensional 
integrals and may be readily evaluated.}
\begin{equation}
K(x,y) \equiv \int_{\Re(w)>0} d^2w \, w^x \wbar^y e^{-2\Re(w)} = 
2\pi^2{\csc[\pi(x+y)]\over\G(-x)\G(-y)}.
\end{equation}
The last equality holds only for $-2 < \Re(x+y) < -1$, but the 
result can be analytically continued outside of this range. This is
necessary in order to evaluate $K$ for physical values of 
$\alpha' t_i$ as well as nonzero $n$ and $m$.

\subsection{Small $\alpha' m_\perp^2$}

As we are interested in ${\cal T}_{26}$ in
the opposite regime $\alpha' m_\perp^2 \ll 1$,
we shall not process the large $\alpha' m_\perp^2$ 
result further. Instead, we make
a change of variables in (\ref{proto}) in order to be able to derive
a power series in $\alpha' m_\perp^2$.
We begin by switching the integral over $u$ in (\ref{proto}) to 
an integral over $uv$. We will see that this change of variables 
misses the saddle point at small $|v|$.  Thus, 
we will eventually need to add a second contribution 
where we switch instead the integral over $v$
to an integral over $uv$.


\begin{FIGURE}[h]{
\centerline{\psfig{figure=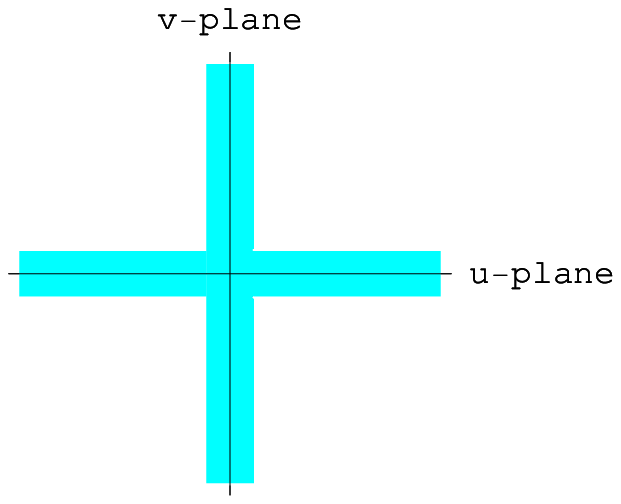}}
\vspace*{-20pt}
\caption{Schematic representation of the dominant integration 
domains for $I$. The vertical strip intersects the $u$-plane 
in a small disc surrounding the origin $(u,\ubar) = (0,0)$, 
while the horizontal strip intersects the $v$-plane in a small 
disc surrounding the origin $(v,\vbar) = (0,0)$.}
\label{fig:saddlepts}
}
\end{FIGURE}


We will call $I_{\rm vert}$ the integral (\ref{proto}) in which we have
replaced an integral over $u$ by an integral over $uv$.
The subscript ``vert'' indicates that we are picking up
the saddle point in the vertical strip in Figure \ref{fig:saddlepts}.
In switching the integral over $u$ to an integral over $uv$, we can 
divide up the domain of integration into the four regions where
$\Re(v)$ and $\Re(uv)$ are either positive or negative,
\begin{equation}
\label{Ivsum}
I_{\rm vert} = \sum I_{\rm vert}^{\sigma_v, \sigma_{uv}} \ .
\end{equation}
We begin by considering the domain where $\Re (v)$ and $\Re (uv)$
are both positive. This necessitates choosing $s_2 = |s_2|e^{i\pi}$ and 
$s = |s|e^{i\pi}$ so that $I_{\rm vert}^{++}$ converges.
Making the change of variables $w = e^{-i\pi}suv$ and 
$\wbar = e^{-i\pi}s\ubar\vbar$, we find
\begin{eqnarray}
& & I_{\rm vert}^{++} = \\
& & (e^{-i\pi}s)^{-2-a-\bar{a}} \int\limits_{\substack{\Re(w) > 0 
\\ \Re(v) > 0}}
d^2w d^2v\, w^a\wbar^{\bar{a}}v^{b-a-1}\vbar^{\bar{b}-\bar{a}-1}
\exp\left[-{s_1\over s}\({w\over v}+{\wbar\over \vbar}\) 
+ s_2(v+\vbar) - (w + \wbar)\right]. \nonumber
\end{eqnarray}
Changing variables to $z = e^{-i\pi}s_2v$ and $\zbar = e^{-i\pi}s_2\vbar$ gives
\begin{equation}
I_{\rm vert}^{++} = (e^{-i\pi}s)^{-2-a-\bar{a}} 
(e^{-i\pi}s_2)^{a+\bar{a}-b-\bar{b}} J_S,
\end{equation}
where
\begin{equation}
\label{J}
J_S \equiv \int\limits_{\substack{\Re(w)>0 \\ \Re(z) > 0}}
d^2w d^2z\, w^a \wbar^{\bar{a}} z^{b-a-1} \zbar^{\bar{b}-\bar{a}-1}
\exp\left[{s_1s_2\over s}\({w\over z} + {\wbar\over \zbar}\) 
- (z + \zbar) - (w + \wbar)\right].
\end{equation}

Next consider the integral $I_{\rm vert}^{+-}$. Now we must choose
$s_2 = |s_2|e^{i\pi}$ and $s = |s|$ so that the integral can
converge. Change variables to $w = suv$ and $\wbar = s\ubar\vbar$, then to 
$z = e^{-i\pi}s_2v$ and $\zbar = e^{-i\pi}s_2\vbar$, and finally rotate 
$w \goesto e^{-i\pi}w$ and $\wbar \goesto e^{i\pi}\wbar$. 
This yields
\begin{equation}
I_{\rm vert}^{+-}  = 
s^{-2-a-\bar{a}} (e^{-i\pi}s_2)^{a+\bar{a}-b-\bar{b}} e^{i\pi(a-\abar)} J_S.
\end{equation}

Thus, for the two terms in (\ref{Ivsum}) with $\Re(v) > 0$ we have found
\begin{equation}
\label{Ivsuma}
I_{\rm vert}^{++} + I_{\rm vert}^{+-} =
\left[(e^{-i\pi}s)^{-2-a-\bar{a}} + e^{i\pi(a-\bar{a})} 
s^{-2-a-\bar{a}}\right](e^{-i\pi}s_2)^{a+\bar{a}-b-\bar{b}} J_S.
\end{equation}
A similar analysis for the four terms with $\Re(v) < 0$ gives
\begin{equation}
\label{Ivsumb}
I_{\rm vert}^{-+} + I_{\rm vert}^{--} = 
\left[e^{i\pi[(b-\bar{b})-(a-\bar{a})]}(e^{-i\pi}s)^{-2-a-\bar{a}} 
+ e^{i\pi(b-\bar{b})} s^{-2-a-\bar{a}}\right]
s_2^{a+\bar{a}-b-\bar{b}} J_S.
\end{equation}
Combining (\ref{Ivsuma}) and (\ref{Ivsumb}),
\begin{equation}
I_{\rm vert} = 
\left[(e^{-i\pi}s)^{-2-a-\bar{a}} + e^{i\pi(a-\bar{a})} s^{-2-a-\bar{a}}\right]
\left[(e^{-i\pi}s_2)^{a+\bar{a}-b-\bar{b}} + 
e^{i\pi[(b-\bar{b})-(a-\bar{a})]} s_2^{a+\bar{a}-b-\bar{b}}\right] J_S.
\end{equation}

Like $J_L$, $J_S$ depends only on the combination $s_1s_2/s$. 
Restoring $\a'$, this is the dimensionless ratio 
$\a'm^2_\perp/4$. As such, we may expand the 
exponentials in (\ref{J}) as a double power series,
\begin{equation}
J_S = \sum_{n,m \geq 0} {(\a'm^2_\perp/4)^{n+m}\over n!m!} 
K(a+n,\abar+m) K(b-a-1-n,\bbar-\abar-1-m) \ .
\end{equation}
Simplifying and using the identity $\pi\csc(\pi z) = \G(z)\G(1-z)$ we obtain
\begin{equation}
\label{Jexplicit}
J_S =  
{4 \pi^2\sin(\pi a)\sin\left[\pi(b-a)\right]\over 
\sin\left[\pi(a+\bar{a})\right]\sin\left[\pi(a+\bar{a}-b-\bar{b})\right]}
\sum_{n,m \geq 0}{(\a'm^2_\perp/4)^{n+m}\over n!m!}
{\G(1+a+n)\G(b-a-n)\over \G(-\bar{a}-m)\G(-\bar{b}+\bar{a}+1+m)}.
\end{equation}

The form of (\ref{J}) along with our power series method of evaluation
makes clear why $I_{\rm vert}$ is not equal to $I$.  We have assumed not just
that $\alpha' m_{\perp}^2 \ll 1$ but that $\alpha' m_\perp^2 w / z \ll 1$
in the domain contributing to the integral.  However, this condition
cuts out the small $z$ region, which in turn corresponds to the small
$v$ region, which is where the other saddle point of $I$
lay.  To cure this problem, we will instead switch in 
(\ref{proto}) the integral over $v$
for an integral over $uv$. This second term $I_{\rm horz}$ will contain
the saddle point at small $v$ but will miss the small $u$ saddle point.
The subscript ``horz'' indicates that we are now picking up contributions
from the horizontal strip in Figure \ref{fig:saddlepts}.
The complete result for $I$ is given by the sum $I_{\rm vert} + I_{\rm horz}$.

Repeating steps (\ref{Ivsum}) to (\ref{Jexplicit}) for 
$I_{\rm horz}$ is straightforward. In fact, $I_{\rm horz}$ can be obtained 
from $I_{\rm vert}$ simply by exchanging $a \leftrightarrow b$, 
$\abar \leftrightarrow \bbar$, $s_2\leftrightarrow s_1$. 
This completes the solution of ${\cal T}_{26}$.

Suppose 
\begin{equation}
\a'm^2_\perp \ll 1
\end{equation}
so that it suffices to keep only the $n=m=0$ term of
(\ref{Jexplicit}). Then $I_{\rm horz}(a+1,a,b,b) = I_{\rm horz}(a,a+1,b,b) = 
(b-a)I_{\rm horz}(a,a,b,b)/s_1$. Also, $I_{\rm vert}(a+1,a,b,b)$ and 
$I_{\rm vert}(a,a+1,b,b)$ are each proportional to 
$(s_2/s)I_{\rm vert}(a,a,b,b)$. When they are combined with the factor of
$s_1$ in (\ref{Tproto}), they contribute at order $\a'm_\perp^2$ which
we are dropping. Thus,
\begin{equation}
{\cal T}_{26}/C \approx -t_2 I_{\rm vert}(a,a,b,b) - t_1 I_{\rm horz}(a,a,b,b).
\end{equation}
This is our main result. Using the identity 
$(e^{-i\pi}z)^x + z^x = 2\(e^{-i\pi/2}z\)^x\cos(\pi x/2)$ and 
restoring factors of $\a'$ gives
\begin{eqnarray}
\label{T26result}
{\cal T}_{26} 
& \approx & 
-\frac{64 \pi^3 g_c^3}{\alpha'} \(e^{-i\pi/2}{\a's\over4}\)^2 \\
& & 
\times \left[~\(e^{-i\pi/2}{\a's\over4}\)^{a't_1/2} 
\(e^{-i\pi/2}{\a's_2\over4}\)^{\a'(t_2-t_1)/2}
\Pi(\a't_1/4,\a'(t_2-t_1)/4)\right.\nonumber\\
& & 
\left. ~~ + \(e^{-i\pi/2}{\a's\over4}\)^{\a't_2/2} 
\(e^{-i\pi/2}{\a's_1\over4}\)^{\a'(t_1-t_2)/2}
\Pi(\a't_2/4,\a'(t_1-t_2)/4)\right] \nonumber
\end{eqnarray}
where $\Pi$ is defined in eq.~(\ref{propagator}).

We can check that this amplitude is consistent with expectations from 
supergravity by expanding the $\Pi$ functions near the graviton poles 
at $t_1 = 0$ and $t_2 = 0$ using eq. (\ref{proplimit}). We find
\begin{eqnarray}
{\cal T}_{26} & \approx & 
\frac{64 \pi^3 g_c^3}{\alpha'} \nonumber \\
& & \times \left[~\,
\(e^{-i\pi/2}{\a's\over4}\)^{2+\a'(t_1+t_2)/4} 
\({s_1\over s_2}\)^{\a'(t_1-t_2)/4} 
{2\sinh\[{\a'\over4}(t_1-t_2)\ln\(e^{-i\pi/2}\a'm_\perp^2/4\)\]\over 
\a'(t_1-t_2)/4}\right. \nonumber \\
& & ~~~~ 
+ {1\over\a't_1/4}\(e^{-i\pi/2}{\a's\over4}\)^{2+\a't_1/2}
\(e^{-i\pi/2}{\a's_2\over4}\)^{\a'(t_2-t_1)/2} \nonumber\\
& & \left. \!~~~~ + {1\over\a't_2/4}
\(e^{-i\pi/2}{\a's\over4}\)^{2+\a't_2/2}
\(e^{-i\pi/2}{\a's_1\over4}\)^{\a'(t_1-t_2)/2} 
\right].
\end{eqnarray}
If $\a't_1$ and $\a't_2$ are slightly less 
than 0, then the first term remains finite, whereas the second and 
third terms diverge and the corresponding Regge exponents for 
$s$ will be slightly less than 2. 

\section{Eigenfunctions in the hard-wall model}
\label{sec:eigfuncs}
Consider the eigenvalue problem $H_i \psi(u) = E \psi(u)$, where $H_i
= -\p_u^2 + 4 - z_0^2 t_i e^{-2u}$ and $E = 4 + \nu^2$ for $\nu >
0$. 
The form of the operator $H_i$ was derived in \cite{Brower:2006ea}.

As a shorthand, denote the dimensionless momentum transfer 
$\rho = z_0\sqrt{|t_i|}$. Then in terms of the variable $\xi =
\rho e^{-u}$, the Schr\"odinger equation is
the modified Bessel differential equation:
\begin{equation}
\xi^2 \psi'' + \xi \psi' - \(\xi^2 + (i\nu)^2\)\psi = 0.
\end{equation}
The general solution 
is in terms of modified Bessel functions of
the first kind,
\begin{equation}
\label{negteigfuncs}
\psi_\nu(u) = c\(I_{i\nu}(\xi) + R(\nu, \rho)  I_{-i\nu}(\xi)\) \ .
\end{equation}

The relative coefficient $R(\nu, \rho) $ is determined by the 
boundary condition at the wall \cite{Brower:2006ea}, 
\begin{equation}
\left.\p_\xi (\xi^2 \psi)\right|_{\xi = \rho} = 0 \ .
\end{equation} 
This boundary condition follows from energy-momentum conservation.
More precisely, note that the metric fluctuation $h_{++} = \xi^{-2}$
is pure gauge because it corresponds to a linear reparametrization
of the background metric.  In order to preserve diffeomorphism
invariance in the bulk and correspondingly energy-momentum
conservation in the boundary gauge theory, we must require
that this pure gauge metric fluctuation satisfy the boundary conditions
at the hard wall.

The boundary condition yields 
\begin{equation}
\label{R}
R(\nu,\rho) = \left.-{\p_\xi(\xi^2 I_{i\nu}(\xi))\over
\p_\xi(\xi^2 I_{-i\nu}(\xi))}\right|_{\xi = \rho}
= -{4I_{i\nu}(\rho) + \rho I_{i\nu-1}(\rho) + \rho I_{i\nu+1}(\rho)
\over 4I_{-i\nu}(\rho) + \rho I_{-i\nu-1}(\rho) + \rho I_{-i\nu+1}(\rho)}.
\end{equation}
Note that at small $\nu$, $R(\nu, \rho) \to -1$.  
This factor of minus one
leads to destructive interference in $\psi_\nu(u)$ near $u=0$.
In section \ref{sec:variousregimes}, our integrals over
$\psi_\nu(u) d\nu$ are dominated by small $\nu$, and thus
this destructive interference has important effects
for the qualitative features of the scattering amplitude. 

The overall coefficient $c$ is fixed by requiring that the eigenfunctions
are delta-function-normalized in the coordinate $u$, 
\begin{equation}
\int_0^\infty d\nu\, \psi^*_\nu(u) \psi_\nu(u') = \d(u-u') \ .
\end{equation}
This gives $|c|^2 = \nu/(2\sinh(\pi\nu))$.\footnote{This normalization 
constant can be obtained by working at 
small momentum transfer $\rho$ where the eigenfunctions look like
plane waves.}
This normalization does not fix the phase of $c$.  We can choose this
phase so that $\psi_\nu(u)$ has a well-defined limit as $\rho \to 0$:
\begin{equation}
\label{c}
c(\nu,\rho) = i\({\rho\over 2}\)^{-i\nu}{\nu\G(i\nu)\over\sqrt{2\pi}} \ .
\end{equation}
Given (\ref{R}) and (\ref{c}), we have solved the eigenvalue problem
$H_i \psi(u) = E \psi(u)$. 

In section \ref{sec:zerot}, we need these eigenfunctions in the 
$t_i=0$ limit.  In this limit of
vanishing momentum transfer, we Taylor expand around $\rho = 0$ to get
\begin{equation}
\label{zeroteigfuncs}
\psi_\nu(u) = 
{1\over\sqrt{2\pi}} \(e^{-i\nu u}+{\nu-2i\over\nu+2i}e^{i\nu u}\) + O(\rho^2).
\end{equation}

\section{Calculation of $\Qbar$}
\label{sec:Qbar}
The integral 
\begin{equation}
\label{qbarinitial}
\Qbar(u,\t_i) = \int_0^\infty du' \phi_0(u')^2 Q(u,u',\t_i)
\end{equation}
may be done by expanding the external hadron wave
functions $\phi_0$ as a power series in $e^{-u'}$
near the UV boundary of our
cavity:
\begin{equation}
\phi_0(u') = e^{-\Delta_0 u'} \sum_{n=0}^\infty c_n e^{-u' n} \ .
\end{equation}
 The
leading term in the expansion will capture the UV behavior of the wave
function and higher order terms will capture the IR
behavior---the details of which depend on the precise physics of 
confinement. 
For our hard-wall model, $\phi_0(u')$ is a Bessel function where
the odd coefficients vanish, $c_{2n+1} = 0$, while the even coefficients
are given by
\begin{equation}
 c_{2n} = {\sqrt{2}\over\La \sqrt{\rm Vol}_W R^{3/2}}
{\({m_0/\La\over2}\)^{2n+\D_0-2}\over J_{\D_0-2}(m_0/\La)} 
\frac{(-1)^n}{n! \, \Gamma(\Delta_0-1+n)}
 \ .
\end{equation}
Inserting the power series for $\phi_0(u')$ in eq.~(\ref{qbarinitial}), 
the resulting expression for $\Qbar$ can be expressed compactly as
\begin{equation}
\label{Qbarformula}
\Qbar(u, \tau_i) = \sum_{n,m \geq 0} c_n c_m 
\int_0^\infty du' e^{-u' (2 \Delta_0 +n+m)} Q(u,u',\tau_i) \ .
\end{equation}
The integral over $Q(u,u',\tau_i)$ was then done exactly using Mathematica 5.2:
\begin{eqnarray}
\label{summand}
\lefteqn{\int_0^\infty du' e^{-2 u' x} Q(u,u',\tau_i) =
\frac{e^{-u^2/4 \tau_i}}{16}
{1\over x-1}\[
f\({-u+4\t_i\over\sqrt{4\t_i}}\)
+ {x^2+3\over x^2-1}f\({u+4\t_i\over\sqrt{4\t_i}}\) \right. } \\
& &~~~~~~~~~~~~~~ \left.
+ 4\sqrt{\t_i}f'\({u+4\t_i\over\sqrt{4\t_i}}\) 
- {2\over x-1}f\({u+4\t_i x\over\sqrt{4\t_i}}\) 
- {2\over x+1}f\({-u+4\t_i x\over\sqrt{4\t_i}}\)
\] \ , \nonumber
\end{eqnarray}
where $f'(y) = -{2\over\sqrt{\pi}} + 2yf(y)$.

We now consider the limit of $\t_i \gg 1$ with $u \ll \tau_i$ in which
case we can obtain beautiful formulae for $\Qbar$, $\Pbar$, and 
${\cal R}(u)$.
In this limit, 
we apply the asymptotic expansion $f(y) \sim {1\over\sqrt{\pi}y}\(1-
{1\over2y^2}+{3 \over 4 y^4}+\ldots\)$ to eq.~(\ref{summand}). 
We find that the integral, and hence the summand in eq.~(\ref{Qbarformula}), 
scales asymptotically as $\t_i^{-3/2}$:
\begin{equation}
\label{tauthreehalves}
\int_0^\infty du' e^{-2 u' x} Q(u,u',\tau_i) \approx 
\frac{(1+2u)}{64 \sqrt{\pi}  \tau_i^{3/2}}
\frac{1+x}{x^2} e^{-u^2/4\tau_i}\ .
\end{equation}
With these formulae and in this limit, we can understand the $\tau_i$ 
and $u$ dependence of $\Qbar$:
\begin{equation}
\label{qbarapprox}
\Qbar(u, \tau_i) \approx
\frac{(1+2u)}{64 \sqrt{\pi}  \tau_i^{3/2}}
 e^{-u^2/4\tau_i}
{\mathcal C} 
\end{equation}
where
\begin{equation}
{\mathcal C} \equiv \sum_{n,m \geq 0} 
\frac{2(2 + 2 \Delta_0 + n + m)}{(2 \Delta_0 + n +m)^2}  c_n c_m  
 \ .
\end{equation}

With knowledge of the $c_n$, ${\mathcal C}$ can be calculated
numerically and rescales $\Qbar$ by some overall constant.  
For example, for our hard-wall
model, in the case $m_0 / \Lambda = 2.405...$ and $\Delta_0=3$, we find that
\begin{equation}
\Qbar(u, \tau_i) \approx  
{2\over\La^2{\rm Vol}_W R^3} \left( 7.270 \times 10^{-3}\right) 
(1+2u)\frac{e^{-u^2/4 \tau_i} }{\tau_i^{3/2}}  
\ .
\end{equation}
(For the hard wall, 
there is in fact an analytic formula for ${\mathcal C}$ as a function 
of $\Delta_0$ and $m_0/\Lambda$
involving ${}_2 F_1$ functions.)

To obtain an expression for ${\cal R}(u)$, recall that
$\Pbar =
4\Qbar - \p_{\t_i}\Qbar$.  In the large $\tau_i$ limit,  
the derivative is subleading, and
we obtain $\Pbar \approx 4 \Qbar$. 
 ${\cal R}(u)$ in 
the large $\tau_i$ limit can then be constructed from (\ref{Rpbarqbar}):
\begin{equation}
{\cal R}(u) \approx {32 \sqrt{\la}\over z_0^4}
\(e^{-i\pi/2}{z_0^2 s\over4\sqrt{\la}}\)^{2-2/\sqrt{\la}}e^{4u/\sqrt{\la}}
\Qbar(u,\t_1)\Qbar(u,\t_2) 
\ .
\end{equation}
Neglecting the $e^{4u / \sqrt{\lambda}}$ because of the large $\lambda$ limit, 
the dependence of this expression for ${\cal R}(u)$ on $u$ and $\tau_i$ is 
\begin{equation}
\label{remarkable}
{\cal R}(u) \sim \frac{(1+2u)^2}{(\tau_1 \tau_2)^{3/2}}
 \exp \[-\frac{u^2}{4} \left(\frac{1}{\tau_1} + \frac{1}{\tau_2}\right)\] 
 \ .
\end{equation}

In order to see how accurately (\ref{remarkable}) approximates
(\ref{Rpbarqbar}), in Figure
\ref{fig:fullratio} we plot the dimensionless version of the double
Pomeron source function, defined in (\ref{dimlessR}), 
for various values of $\t_1$ and $\t_2$. As
expected, the agreement gets better as each $\t_i$ increases.

\begin{FIGURE}[h]{
\centerline{
a) \psfig{figure=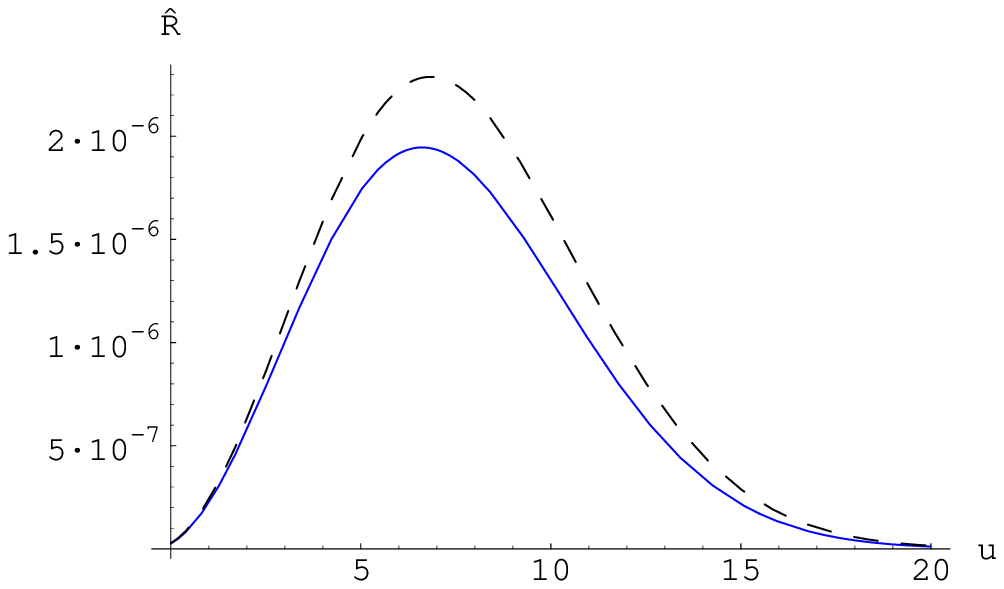,width=2.5in} 
b) \psfig{figure=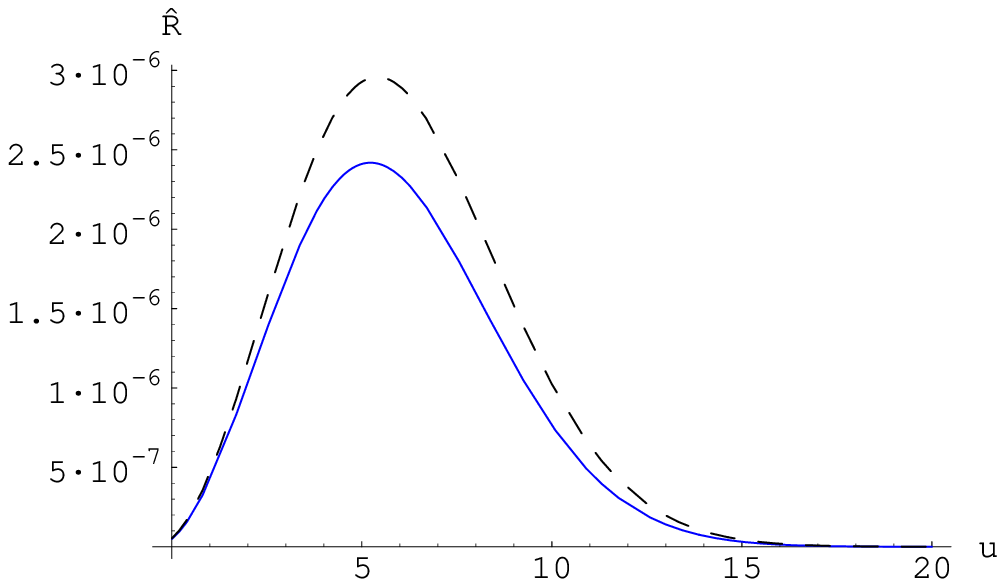,width=2.5in}}
\centerline{c) \psfig{figure=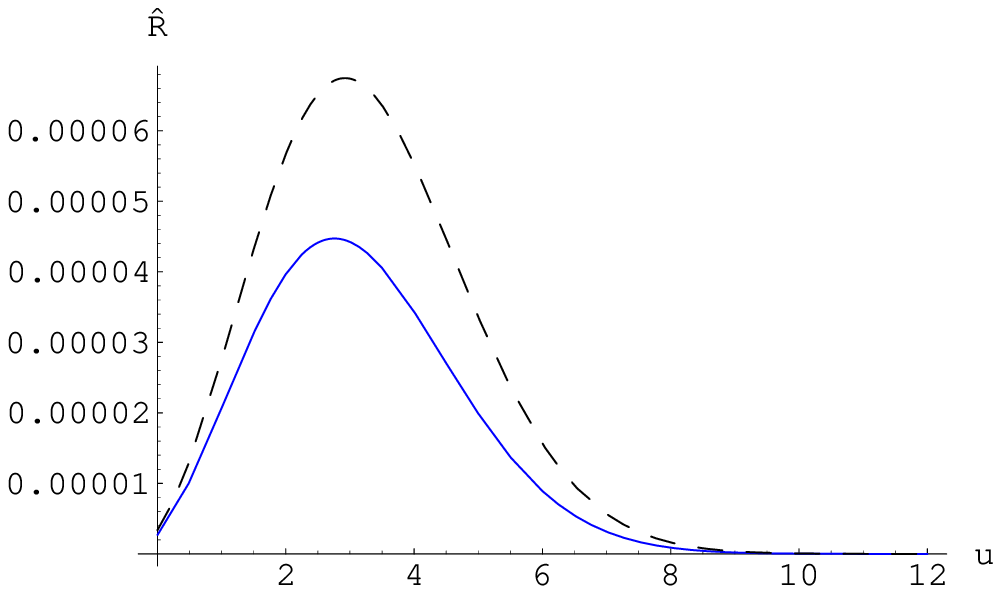,width=2.5in}}
\vspace*{-20pt}
\caption{We plot $\widehat {\mathcal R}(u)$ for various values of
$\tau_1$ and $\tau_2$.  The solid curve is the exact numerical
value of $\widehat {\mathcal R}(u)$ as a function of $u$.  The dashed
curve is our approximate formula (\ref{remarkable}).  We have
plotted a) $\tau_1 = \tau_2 = 25$, b) $\tau_1 = 10$, $\tau_2=40$, and
c) $\tau_1=\tau_2=5$.  The approximation, always too large, improves
as $\tau_i$ increases.}
\label{fig:fullratio}
}
\end{FIGURE}

We also examine the accuracy of eq.~(\ref{Rmax})
by plotting in Figure \ref{fig:ratiocompare} 
its predicted value for ${\mathcal R}(u_{\rm max})/{\mathcal R}(0)$ 
against the actual value obtained numerically from (\ref{Rpbarqbar}).

\begin{FIGURE}[h]{
\centerline{\psfig{figure=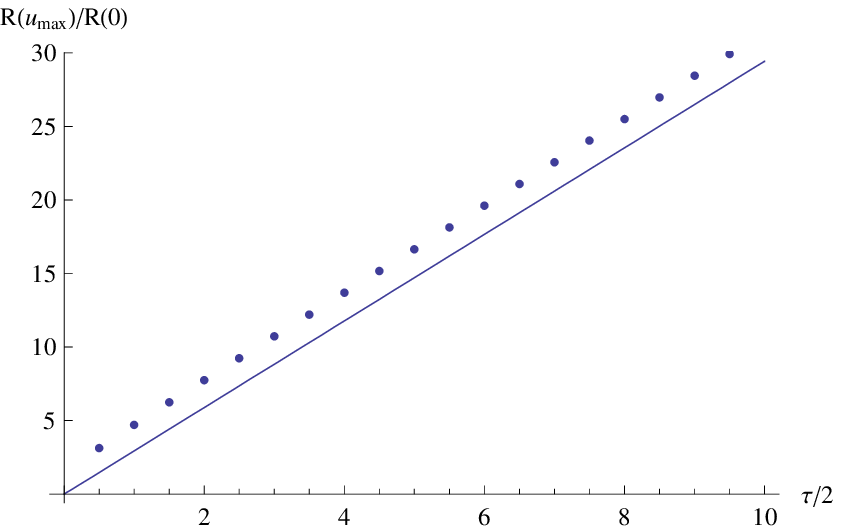,width=2.5in}}
\vspace*{-20pt}
\caption{
We plot the ratio 
${\mathcal R}(u_{\rm max}) / {\mathcal R}(0)$ 
as a function of $\tau/2$ with $\tau_1 = \tau_2$.  The points are the 
numerically determined values of the ratio, while the solid line is
the approximation eq.~(\ref{Rmax}).
}
\label{fig:ratiocompare}
}
\end{FIGURE}

\end{document}